\documentclass[10pt]{article}
\usepackage[utf8x]{inputenc}
\usepackage[margin=1in]{geometry}
\usepackage{amssymb,amsmath,amsthm,amsfonts}
\usepackage{mathtools}
\usepackage{hyperref}
\usepackage{bbm}

\usepackage{natbib}
\usepackage{graphicx}
\usepackage{verbatim}
\usepackage{enumerate}
\usepackage{multirow}
\usepackage{subcaption}

\graphicspath{{Images/}}

\newcommand{\crps}{\mathrm{CRPS}}
\newcommand{\one}{\mathbbm{1}}
\newcommand{\dd}{\: \mathrm{d}}
\newcommand{\R}{\mathbb{R}}

\newtheorem{propo}{Proposition}
\newtheorem{theorem}{Theorem}

\theoremstyle{definition}
\newtheorem{defin}{Definition}
\newtheorem{remark}{Remark}

\title{Evaluating forecasts for high-impact events using transformed kernel scores}
\author{Sam Allen \qquad David Ginsbourger \qquad Johanna Ziegel \\[1ex]{\normalsize Institute of Mathematical Statistics and Actuarial Science}\\ {\normalsize University of Bern}\\{\normalsize Bern, Switzerland}\\ {\normalsize\url{{sam.allen,david.ginsbourger,johanna.ziegel}@stat.unibe.ch}}}
\date{}

\begin{document}

\maketitle

\begin{abstract}
    It is informative to evaluate a forecaster's ability to predict outcomes that have a large impact on the forecast user. Although weighted scoring rules have become a well-established tool to achieve this, such scores have been studied almost exclusively in the univariate case, with interest typically placed on extreme events. However, a large impact may also result from events not considered to be extreme from a statistical perspective: the interaction of several moderate events could also generate a high impact. Compound weather events provide a good example of this. To assess forecasts made for high-impact events, this work extends existing results on weighted scoring rules by introducing weighted multivariate scores. To do so, we utilise kernel scores. We demonstrate that the threshold-weighted continuous ranked probability score (twCRPS), arguably the most well-known weighted scoring rule, is a kernel score. This result leads to a convenient representation of the twCRPS when the forecast is an ensemble, and also permits a generalisation that can be employed with alternative kernels, allowing us to introduce, for example, a threshold-weighted energy score and threshold-weighted variogram score. To illustrate the additional information that these weighted multivariate scoring rules provide, results are presented for a case study in which the weighted scores are used to evaluate daily precipitation accumulation forecasts, with particular interest on events that could lead to flooding.
\end{abstract}

\section{Introduction}
\label{section:Intro}



Just as important as issuing a forecast is understanding how it is expected to perform. In achieving this, forecasters gain a greater awareness of the strengths and limitations of their predictions, and, in turn, learn how they can be improved \citep{Jolliffe2012}. There are several aspects to consider when evaluating forecasts, but, intuitively, a `good' forecaster is one whose predictions consistently agree with what materialises \citep{Murphy1993}. To assess to what extent this is satisfied, it is convenient to condense all information regarding forecast performance into a single numerical value, or score, thereby allowing competing forecast strategies to be objectively ranked and compared. For probabilistic forecasts, this can be achieved using scoring rules. Scoring rules are functions of the form
\begin{equation*}
    S \colon \mathcal{M} \times \mathcal{X} \to \mathbb{R} \cup \{ -\infty, \infty \},
\end{equation*}
where $ \mathcal{M} $ is a suitable class of probability measures over the measurable outcome space ($ \mathcal{X} , \mathcal{A} $). Let $ \mathfrak{M} $ denote the set of all probability measures on ($ \mathcal{X} , \mathcal{A} $). Without loss of generality, we restrict attention to negatively oriented scoring rules, for which a lower score indicates a more accurate forecast. The score assigned to a forecast can therefore be interpreted as a loss.

\bigskip


It is widely accepted that scoring rules should be proper. A scoring rule $ S $ is proper with respect to $ \mathcal{M} \subset \mathfrak{M} $ if 
\begin{equation}\label{eq:Proper}
    S(Q, Q) \leq S(P, Q) \hspace{1cm} \textrm{for all} \hspace{0.1cm} P, Q \in \mathcal{M},
\end{equation}
where $ S(P, Q) = \mathbb{E}_{Q}[S(P, Y)] $ denotes the expectation of $ S(P, Y) $ when $ Y \sim Q $; it is assumed that $ S(P, Q) $ exists for all $ P,Q \in \mathcal{M} $, and that $ S(Q, Q) $ is finite. That is, if the observations are believed to arise according to $ Q \in \mathcal{M} $, then the expected value of a proper score is minimised by issuing $ Q $ as the forecast. If $ Q $ is the unique minimiser of the expected score, then the scoring rule is said to be strictly proper.

\bigskip

It is often of interest to evaluate a forecaster's ability to predict outcomes that have a large impact on the user, since improving the forecasts made for such outcomes may allow their impacts to be mitigated. However, \cite{GneitingRanjan2011} demonstrate that using a proper scoring rule to evaluate the predictions made when particular outcomes occur is equivalent to assessing the forecaster using an improper score, which can thus result in unreliable conclusions regarding the performance of competing forecasters. In the context of extreme events, \cite{Lerch2017} term this the forecaster's dilemma. To circumvent the forecaster's dilemma, it has become common to employ weighted scoring rules that can direct the evaluation of forecasts to certain outcomes in a theoretically sound way.

\bigskip


The concept of weighted scoring rules dates back at least to \cite{Matheson1976}, though most developments in the field have occurred over the past decade. \cite{GneitingRanjan2011}, for example, introduce two weighted versions of the continuous ranked probability score (CRPS), while \cite{Diks2011} propose two adaptations of the logarithmic score. \cite{Holzmann2017} generalise the approach followed by \cite{Diks2011} and present a framework capable of constructing weighted versions of any proper scoring rule. In order to focus evaluation on particular outcomes, in this paper, a weight function is a measurable function from the sample space $ \mathcal{X} $ to $ [0, 1] $. Alternative weight functions have also been considered in the literature \citep[see e.g.][]{GneitingRanjan2011}, but they are not of interest here. 

\bigskip


Weighted scoring rules have been studied in most detail in the univariate setting, with interest often placed on extreme events, defined as instances where the outcome exceeds a chosen threshold. Often, however, a high-impact results from the interaction of several moderate events, none of which are extreme from a statistical perspective. A good example of this is a compound weather event, whereby multiple weather hazards combine and interact to generate a high-impact event, despite none of the confounding hazards themselves necessarily being `extreme' \citep[e.g.][]{Zscheischler2020}. The present article seeks to develop tools that permit a targeted assessment of forecasts made for high-impact events such as these. In particular, we introduce weighted multivariate scoring rules that allow emphasis to be placed on regions of a multi-dimensional outcome space when evaluating the accuracy of a forecaster. 

\bigskip


To achieve this, we utilise kernel scores, a general class of proper scoring rules based on conditionally negative definite (c.n.d.) kernels \citep{GneitingRaftery2007, Dawid2007}. Here, a negative definite kernel is a symmetric function $ \rho: \mathcal{X} \times \mathcal{X} \to \R $ for which
\begin{equation}\label{eq:NDKernel}
    \sum_{i=1}^{n}\sum_{j=1}^{n} c_{i} c_{j} \rho(x_{i}, x_{j}) \leq 0
\end{equation}
for all $ n \in \mathbb{N}, x_{1}, \dots, x_{n} \in \mathcal{X}, $ and $ c_{1}, \dots, c_{n} \in \R $. A kernel is c.n.d.~if the above criterion is satisfied for all $ c_{1}, \dots, c_{n} \in \R $ that sum to zero, and is strictly negative definite if equality in Equation \ref{eq:NDKernel} holds only when $ c_{1} = 0, \dots, c_{n} = 0 $ for distinct $ x_{1}, \dots, x_{n} $. Conversely, a kernel is said to be positive definite if the inequality in Equation \ref{eq:NDKernel} is reversed. Hereinafter, we assume wherever necessary that the kernels are measurable.

\begin{defin}
    Given a c.n.d.~kernel $ \rho: \mathcal{X} \times \mathcal{X} \to \R $, the \emph{kernel score corresponding to $ \rho $} is the scoring rule 
    \begin{equation}\label{eq:KernelS}
        S_{\rho}(P, y) = \mathbb{E}_{P}[\rho(X, y)] - \frac{1}{2}\mathbb{E}_{P}[\rho(X, X^{\prime})] - \frac{1}{2}\rho(y, y),
    \end{equation}
    where $ X, X^{\prime} \sim P \in \mathfrak{M} $ are independent, and it is assumed that all expectations are finite.
\end{defin}

Several familiar scoring rules fall into this kernel score framework, including the Brier score \citep{Brier1950}, the CRPS, and the energy score \citep{GneitingRaftery2007}. In Section \ref{section:wES}, we demonstrate that the variogram score proposed by \cite{Scheuerer2015} is also a kernel score. The final term in Equation \ref{eq:KernelS} does not depend on the forecast and is not present in previous definitions of kernel scores \citep{Gneiting2007, Steinwart2019}. Nonetheless, it is included here since it generates a scoring rule that can be interpreted as a divergence between the forecast and a Dirac measure at the outcome, even if $\rho(y, y) \neq 0 $. 

\bigskip

The deployment of c.n.d.~kernels within scoring rules follows from their interpretation as generalised distance measures \citep[e.g.][]{Scholkopf2001}. However, the propriety of a kernel score depends on the choice of $ \rho $. To see this, consider the divergence function $ d(P, Q) = S(P, Q) - S(Q, Q) $ associated with the scoring rule $ S $. It is immediate from Equation \ref{eq:Proper} that a scoring rule is proper with respect to $ \mathcal{M} $ if and only if its divergence function is non-negative for all $ P,Q \in \mathcal{M} $. The divergence function corresponding to a kernel score is
\begin{equation}\label{eq:KernelSDiv}
   d_{\rho}(P, Q) = \mathbb{E}_{P,Q} \left[ \rho(X, Y) \right] - \frac{1}{2} \mathbb{E}_{P} \left[ \rho(X, X^{\prime}) \right] - \frac{1}{2} \mathbb{E}_{Q} \left[ \rho(Y, Y^{\prime}) \right],
\end{equation}
where $ X, X^{\prime} \sim P $ and $ Y, Y^{\prime} \sim Q $ are independent. That is, the score divergence between $ P $ and $ Q $ is proportional (by a factor of one half) to the energy distance with respect to $ \rho $ \citep{Szekely2013}. \cite{Sejdinovic2013} show that energy distances are special cases of squared Maximum Mean Discrepancies \citep[MMD;][]{Gretton2007}, and kernel score divergences can be interpreted as squared MMDs under suitable integrability conditions. This provides a natural connection between the optimum score estimation theory introduced by \cite{GneitingRaftery2007} and machine learning algorithms that use the MMD as a loss function \citep[e.g.][]{Dziugaite2015, Li2015}. 

\bigskip

The following theorem summarises existing results on the propriety of kernel scores, and is obtained by merging results in \cite{Sejdinovic2013} and \cite{Steinwart2019}. Previously, a special case of this result was presented in \cite{GneitingRaftery2007}. 

\begin{theorem}\label{theorem:Proper}
Let $\rho$ be a c.n.d.~kernel on $\mathcal{X}$. If $\rho(x, x)=0$ for all $x \in \mathcal{X}$, then $\rho$ is non-negative and the kernel score $S_{\rho}$ is proper with respect to 
$$ \mathcal{M}_{\rho} = \{ P \in \mathfrak{M} \: | \: \mathbb{E}_{P}[\rho(X, x_{0})] < \infty \: \text{for some} \: x_{0} \in \mathcal{X} \}. $$
If $\rho$ is negative definite, then the kernel score $S_{\rho}$ is proper with respect to 
$$ \mathcal{M}^{\rho} = \{ P \in \mathfrak{M} \: | \: \mathbb{E}_{P}\left[ \sqrt{-\rho(X, X)} \right] < \infty \}. $$
\end{theorem}

For a c.n.d.~kernel $\rho$, we will often state the assumption that $S_\rho$ is proper with respect to $\mathcal{M}_\rho$ or $\mathcal{M}^\rho$. For concision, it is always assumed in the former case that $\rho(x, x)=0$ for all $x \in \mathcal{X}$, and in the latter case that $\rho$ is negative definite.

\bigskip

The strict propriety of kernel scores relates to injectivity of kernel mean embeddings \citep{Steinwart2019}. If $\rho = -k $, with $k$ being a positive definite and bounded kernel, then this is synonymous with the kernel being characteristic \citep{Muandet2017}. On the other hand, if the c.n.d.~kernel $\rho$ is a metric, then the criterion that Equation \ref{eq:KernelSDiv} is zero if and only if $ P = Q $ is exactly the definition given by \cite{Lyons2013} for $ \rho $ to be a metric of strong negative type. We will refer to specific results where necessary throughout the paper.

\bigskip


Since the theory underlying kernels is well-established, results from the extant literature can be leveraged in order to choose the most appropriate kernel when evaluating forecasts in particular scenarios. The kernel can be chosen to extract the information that is most relevant for the situation at hand, allowing prior information to be incorporated directly into forecast evaluation. \cite{Bolin2019}, for example, propose altering the kernel used within the CRPS to reduce the score's sensitivity to outliers. We study the choice of kernel in the context of weighted scoring rules, with particular interest on forecasts made for high-impact events.

\bigskip


In the following section, we demonstrate that the threshold-weighted continuous ranked probability score (twCRPS) introduced by \cite{GneitingRanjan2011} is a kernel score. The twCRPS is arguably the most well-known weighted scoring rule, and this result leads to a convenient representation of the score when evaluating ensemble forecasts, i.e.~finite samples of point forecasts. In addition, this permits a generalisation of the twCRPS to so-called threshold-weighted kernel scores, which we introduce in Section \ref{section:MNKSFO}. Furthermore, established results on kernels are leveraged in order to introduce further new approaches to weighting kernel scores. Due to the flexibility of kernels, these results significantly widen the range of situations in which weighted scoring rules can be applied, and we illustrate this in Section \ref{section:wES} by considering outcomes in multi-dimensional Euclidean space. We introduce weighted variogram scores and weighted energy scores, and also study a new scoring rule based on a bounded kernel. The utility of these weighted multivariate scoring rules when evaluating forecasts made for high-impact events is presented in a simulation study, as well as in a case study on flood forecasts in Section \ref{section:CaseStudy}. A discussion of the results presented herein is available in Section \ref{section:Discussion}, while all proofs are deferred to the appendix.

\section{Weighted versions of the CRPS}
\label{section:wCRPS}

\subsection{Definitions and properties}

Here, we consider the case where $ \mathcal{M} $ is the set of Borel probability measures on $ \mathcal{X} = \R $ with finite first moment, and identify elements of $ \mathcal{M} $ with their associated distribution functions. A popular scoring rule used to assess forecasts in this setting is the continuous ranked probability score (CRPS), defined as 
\begin{equation}
\label{eq:crps}
\begin{split}
    \crps(F, y) & = \int_{\R} (F(z) - \one\{y\leq z\})^2 \dd z, \\
    & = 2\int_{(0,1)} (\one\{F^{-1}(\alpha) \geq y\} - \alpha)(F^{-1}(\alpha) - y) \dd \alpha, \\
    & = \mathbb{E}_{F}|X - y| - \frac{1}{2}\mathbb{E}_{F}|X - X^{\prime}|,
\end{split}
\end{equation}
where $ X $ and $ X^{\prime} $ are independent random variables with distribution function $ F \in \mathcal{M}$, $ y \in \R $ is the corresponding observation, and $ \one $ denotes the indicator function. In the second expression, $ F^{-1} $ is the lower quantile function or generalised inverse of $ F $. 

\bigskip

The CRPS is strictly proper with respect to $ \mathcal{M} $ \citep{GneitingRaftery2007}. Its three different representations also partly explain the score's popularity. The first expression demonstrates that the CRPS is equivalent to the Brier score integrated over all possible thresholds \citep{Matheson1976}, whereas the second expression highlights that it can also be written as a quantile scoring rule integrated over all quantiles \citep{Laio2007}. The final representation demonstrates that the CRPS is a kernel score, where the c.n.d.~kernel is $\rho(x, x^{\prime}) = | x - x^{\prime} | $ \citep{GneitingRaftery2007}.

\bigskip

Due to its popularity, weighted scoring rules have been studied in most detail using the CRPS, with the most well-known version being the threshold-weighted continuous ranked probability score (twCRPS):
\begin{equation}
\label{eq:twcrps}
    \mathrm{twCRPS}(F, y; \nu ) = \int_{\R} (F(z) - \one\{y\leq z\})^2 \dd \nu(z),
\end{equation}
where $ \nu $ is a Borel measure on $ \R $, often chosen so that it has density equal to a particular non-negative weight function, $ w $ \citep{Matheson1976, GneitingRanjan2011}. Although analytical expressions of the twCRPS have been derived for particular families of parametric distributions \citep[e.g.][]{Allen2021}, the integral in Equation \ref{eq:twcrps} is often evaluated using numerical techniques. The following proposition provides an alternative representation of the twCRPS as a kernel score, implying a straightforward approach to computing this integral when $ F $ is an empirical distribution function.

\begin{propo}\label{prop:twcrps}
Let $ \nu $ be a Borel measure on $ \R $. Then, there exists an increasing function $ v $ on $ \R $ such that the threshold-weighted CRPS associated with the measure $ \nu $ is the kernel score corresponding to $ \rho(x, x^{\prime})  = |v(x) - v(x^{\prime})| $. In particular, $ v $ is any function such that $ v(x) - v(x^{\prime}) = \nu([x^{\prime}, x)) $ for all $ x, x^{\prime} \in \R $. For $ F \in \mathcal{M}_{\rho} $, it holds that
\begin{equation}\label{eq:twcrps2}
\begin{split}
    \mathrm{twCRPS}(F, y; \nu) & = \int_{\R} (F(z) - \one\{y\leq z\})^2 \dd \nu(z), \\
    & = 2\int_{(0,1)} (\one\{F^{-1}(\alpha) \geq y\} - \alpha)(v(F^{-1}(\alpha)) - v(y)) \dd \alpha, \\
    & = \mathbb{E}_{F} \vert v(X) - v(y) \vert - \frac{1}{2}\mathbb{E}_{F} \vert v(X) - v(X^{\prime}) \vert,
\end{split}
\end{equation}
where $ X, X^{\prime} \sim F $ are independent and $ y \in \R $.
\end{propo}

\bigskip

Equation \ref{eq:twcrps2} generalizes the three representations of the CRPS in Equation \ref{eq:crps}, which correspond to the case where $ \nu $ is the Lebesgue measure and $ v(z) = z $ (up to a constant) for all $ z \in \R $. The final equality in Proposition \ref{prop:twcrps} illustrates that the twCRPS can be interpreted as the CRPS after having deformed the forecasts and observations, where the deformation is governed by the choice of measure, or weight function. We refer to $ v $ as the chaining function. 

\begin{remark}\label{remark:twcrps}
    If the measure $ \nu $ in the threshold-weighted CRPS is chosen so that it has density $ w $, then the chaining function $ v $ is any function such that 
    $$ v(x) - v(x^{\prime}) = \int_{[x^{\prime}, x)} w(x) \dd x. $$
\end{remark}

\begin{figure}[!t]
    \centering
    \includegraphics[width=0.3\linewidth]{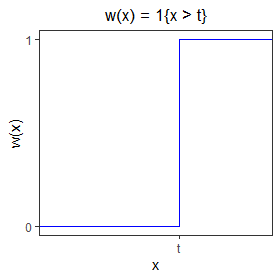}
    \includegraphics[width=0.3\linewidth]{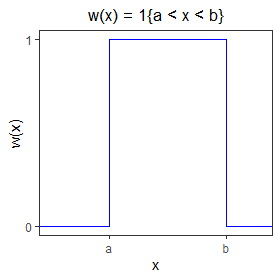}
    \includegraphics[width=0.3\linewidth]{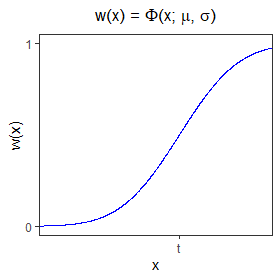}
    \includegraphics[width=0.3\linewidth]{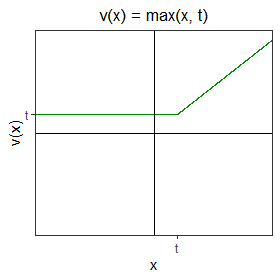}
    \includegraphics[width=0.3\linewidth]{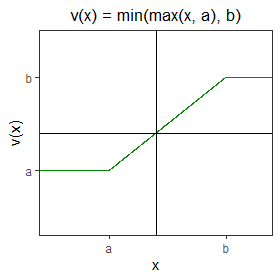}
    \includegraphics[width=0.3\linewidth]{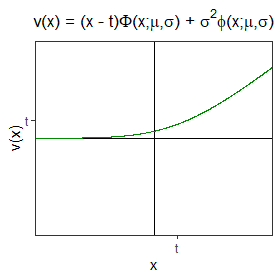}
    \includegraphics[width=0.3\linewidth]{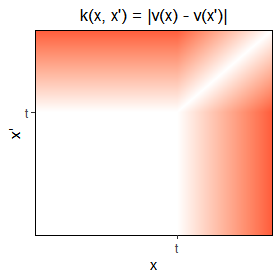}
    \includegraphics[width=0.3\linewidth]{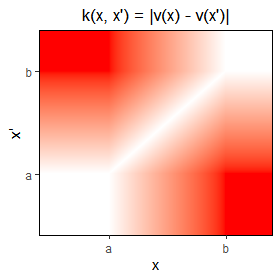}
    \includegraphics[width=0.3\linewidth]{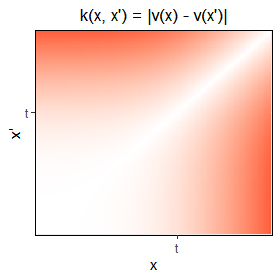}
    \caption{Common weight functions (top row) with corresponding chaining functions (middle row) and the resulting kernels to be employed in the twCRPS (bottom row). $ \Phi(x; \mu, \sigma) $ and $ \phi(x; \mu, \sigma) $ represent the distribution and density functions, respectively, of a normal distribution with mean $ \mu $ and scale $ \sigma $. In the bottom row, a stronger shade of red reflects a larger value of the kernel.}
    \label{fig:Deformations}
\end{figure}

In light of this remark, possible deformations corresponding to some common weight functions are displayed in Figure \ref{fig:Deformations}, along with the resulting kernel to be employed in the twCRPS. Knowing the chaining function permits a greater appreciation of what the weight in the twCRPS achieves. For example, if weight is placed only on values above a certain threshold, $ w(z) = \one\{z \geq t\} $, then the forecasts and observations are projected onto $ [t, \infty) $, with values lower than the threshold mapped onto $ t $, before calculating the unweighted CRPS.

\begin{remark}
    The kernel score representation of the twCRPS readily extends to the integrated quadratic distance (IQD), the score divergence associated with the CRPS \citep{Thorarinsdottir2013}. A threshold-weighted version of the IQD is defined analogously to the twCRPS in Equation \ref{eq:twcrps}, and there exists a similar representation of this weighted divergence in terms of the kernel $ \rho(x, x^{\prime}) = |v(x) - v(x^{\prime})| $.
\end{remark}

\bigskip

The second equality in Proposition \ref{prop:twcrps} demonstrates that, like the CRPS, the twCRPS can also be expressed as the integral of a quantile scoring rule over all possible quantiles. Any scoring function that is consistent for the $ \alpha $-quantile can be written in the form $(\one\{x \geq y\} - \alpha)(v(x) - v(y)) $, with $ v $ increasing \citep{Gneiting2011}, and we define a quantile scoring rule by replacing $ x $ with the $ \alpha $-quantile of a forecast distribution. Since any increasing function $ v $ satisfies $ v(x) - v(x^{\prime}) = \nu([x^{\prime}, x)) $ for some Borel measure $ \nu $, Proposition \ref{prop:twcrps} shows that the integral of any scoring rule that is consistent for the $ \alpha $-quantile over all possible values of $ \alpha $ results in a twCRPS. \cite{GneitingRanjan2011} use the quantile score representation of the CRPS to introduce a quantile-weighted version of the CRPS. In this case, the integral is over the quantiles of the forecast distribution, and the weight function emphasises particular regions of the forecast distribution, rather than regions of the outcome space. Since the threshold-weighted CRPS can also be expressed as the integral of a quantile scoring rule, it would be straightforward to introduce a weight function into this integral in an analogous way, thereby constructing a weighted version of the CRPS that emphasises certain regions of both the outcome space and the forecast distribution.

\bigskip

The threshold- and quantile-weighted versions of the CRPS were introduced to circumvent the fact that scaling a proper scoring rule using a weight governed by the outcomes results in an improper scoring rule \citep{GneitingRanjan2011}. Let $ S $ denote a scoring rule that is proper with respect to $ \mathcal{M} $, let $ w $ be a weight function, and let $ G \in \mathcal{M} $ be a distribution function for which $ \mathbb{E}_{G}[w(X)] > 0 $. The expectation of the score $ S $ scaled by the weight function $ w $, $ \mathbb{E}_{G}[w(Y)S(F, Y)] $, is minimised by issuing a weighted version of $ G $, rather than $ G $ itself. In particular, the expected scaled score is minimised by
\begin{equation}
\label{eq:weightedG}
    G_{w}(x) = \frac{\mathbb{E}_{G}\left[ \one\{ X \leq x\} w(X) \right] }{\mathbb{E}_{G}[w(X)]}.
\end{equation}
\cite{Holzmann2017} therefore suggest assessing forecast distributions through their weighted representation: if $ w $ is a weight function and $ S $ is a proper scoring rule with respect to $\{ F_{w} | F \in \mathcal{M} , \mathbb{E}_{F}[w(X)] > 0\} $, with $ F_{w} $ defined as in Equation \ref{eq:weightedG}, then
\begin{equation}\label{eq:HK17}
    \mathrm{ow}S(F, y; w) = w(y) S(F_{w}, y)
\end{equation} 
defines a scoring rule that is proper with respect to $ \{ F \in \mathcal{M} | \mathbb{E}_{F}[w(X)] > 0 \}$. Since the weighting in this case is directly dependent on the outcome $ y $, we refer to scoring rules in this form as outcome-weighted scoring rules. For example, the outcome-weighted CRPS is defined as
\begin{equation*}
    \mathrm{owCRPS}(F, y; w) = w(y)\int_{\R} (F_{w}(z) - \one\{y \leq z\})^{2} \dd z.
\end{equation*}

\bigskip

To appreciate how the outcome-weighted CRPS differs from the threshold-weighted CRPS, consider a weight function of the form $ w(z) = \one\{z \geq t\} $. The twCRPS with this weight function assesses to what extent the forecast can identify whether or not the observation will exceed the threshold $ t $ or any value larger than $ t $. The owCRPS, on the other hand, is concerned with the forecast performance only when the observation $ y $ exceeds this threshold, in which case the forecast is assessed via its conditional distribution given that the threshold has been exceeded. Alternatively, this difference is made more explicit by noting that the outcome-weighted CRPS can be expressed in the following form.

\begin{propo}\label{prop:owcrps}
Let $ w $ be a weight function. If $ F \in \mathcal{M} $ such that $ \mathbb{E}_{F}[w(X)] > 0 $, then it holds that
\begin{equation}
\label{eq:owcrps2}
    \mathrm{owCRPS}(F, y; w) = \frac{1}{C_{w}(F)}\mathbb{E}_{F} \left[ |X - y|w(X)w(y) \right] - \frac{1}{2C_{w}(F)^{2}}\mathbb{E}_{F} \left[ |X - X^{\prime}|w(X)w(X^{\prime})w(y) \right],
\end{equation}
where $ X, X^{\prime} \sim F $ are independent, $ y \in \R $, and $ C_{w}(F) = \mathbb{E}_{F}[w(X)] $.
\end{propo}

\cite{Holzmann2017} provide this expression for weight functions of the form $ w(z) = \one\{z \geq t\} $. This representation of the outcome-weighted CRPS demonstrates that the weight function is applied to the output of the kernel, in contrast to the threshold-weighted version of the CRPS, which involves a prior transformation of the forecasts and outcome. 

\bigskip

Note also that, unlike the twCRPS, the owCRPS is not a kernel score. To see this, consider a forecast that is a Dirac measure at $ z \in \R $, $ P = \delta_{z} $. From Equation \ref{eq:KernelS}, it is straightforward to verify that a kernel score $ S_{\rho} $ satisfies
\begin{equation*}
    S_{\rho}(\delta_{z}, y) = \rho(z, y) - \frac{1}{2}\rho(z, z) - \frac{1}{2}\rho(y, y) = S_{\rho}(\delta_{y}, z).
\end{equation*}
However, assuming $ w(y), w(z) > 0 $, $ \mathrm{owCRPS}(\delta_{z}, y; w) = |z - y|w(y) $, which is in general not equal to $ \mathrm{owCRPS}(\delta_{y}, z; w) = |z - y|w(z) $.

\subsection{Localising scores}
\label{section:LocalProp}

\cite{Holzmann2017} show that outcome-weighted scoring rules exhibit some desirable properties when interest is on only a particular subset of outcomes. We recall their definitions of localising, and strictly or proportionally locally proper scoring rules.

\begin{defin}
    Let $\mathcal{M}$ be a class of probability measures on a measurable space $(\mathcal{X}, \mathcal{A})$, $ S $ a scoring rule, and $ w $ a weight function. The scoring rule $S$ is called \emph{localising with respect to $ w $} if, for any $ P, Q \in \mathcal{M} $, $ P(\cdot \cap \{ w > 0\} ) = Q( \cdot \cap \{w > 0\} ) $ implies that $ S(P, y) = S(Q, y) $ for all $ y \in \mathcal{X} $. Here, $ \{w > 0\} = \{x \in \mathcal{X} | w(x) > 0\} $.
\end{defin}

That is, a weighted scoring rule is localising if it depends only on the forecast measure restricted to outcomes for which the weight function is positive. Clearly, however, if $ \{ w = 0 \} = \{x \in \mathcal{X} | w(x) = 0\}$ is non-empty, then such a scoring rule will not be strictly proper with respect to typical choices of $ \mathcal{M} $.

\begin{defin}
    Let $S$ be a scoring rule that is proper with respect to $ \mathcal{M}$ and localising with respect to $w$. Then, $S$ is called \emph{strictly locally proper with respect to $w$ and $ \mathcal{M}$} if $ S(P, Q) = S(Q, Q) $ implies $ P(\cdot \cap \{ w > 0\} ) = Q( \cdot \cap \{w > 0\} ) $ for any $ P, Q \in \mathcal{M} $. The scoring rule $S$ is called \emph{proportionally locally proper with respect to $w$ and $ \mathcal{M}$} if $ S(P, Q) = S(Q, Q) $ holds if and only if $ P(\cdot \cap \{ w > 0\} ) = cQ( \cdot \cap \{w > 0\} ) $ for some constant $ c > 0 $ that depends on $ P$ and $ Q$, for any $P, Q \in \mathcal{M}$.
\end{defin}

If $ S $ is a proper scoring rule with respect to $\{ P_{w} | P \in \mathcal{M} , \mathbb{E}_{P}[w(X)] > 0\} $, \citet[][Theorem 3]{Holzmann2017} prove that the outcome-weighted version of this score (constructed via Equation \ref{eq:HK17}) will be proper and localising with respect to $ \{ P \in \mathcal{M} | \mathbb{E}_{P}[w(X)] > 0\}$, and if $ S $ is strictly proper, then the outcome-weighted score will additionally be proportionally locally proper. In order to obtain a strictly locally proper weighted scoring rule, the authors suggested complementing an outcome-weighted score $ S $ with a strictly proper scoring rule $ S_{0} $ for probability forecasts of the occurrence of an arbitrary binary event, such as the logarithmic score or the Brier score:
\begin{equation}\label{eq:owComp}
    \tilde{S}(P, y) = S(P, y) + \{ w(y) S_{0}(C_{w}(P), 1) + (1 - w(y)) S_{0}(C_{w}(P), 0) \},
\end{equation}
where $ C_{w}(P) $ is as defined in Proposition \ref{prop:owcrps}, and is interpreted here as a probability forecast. If $ S $ is proportionally locally proper, then $ \tilde{S} $ will be strictly locally proper \citep[][Theorem 3]{Holzmann2017}. To understand why this holds, note that proportionally locally proper scores only evaluate the shape of the restriction of $ P $ to $ \{ w > 0 \} $, whereas the binary score included in Equation \ref{eq:owComp} allows $ \tilde{S} $ to additionally assess the measure that is assigned to the sets $ \{ w = 0 \} $ and $ \{ w > 0 \} $.

\bigskip 

While outcome-weighted scoring rules are localising by construction, the twCRPS is only localising with respect to particular weight functions. The twCRPS is not localising with respect to weights that are the indicator function of a compact interval over the real line, $ w(z) = \one\{a \leq z \leq b\} $, whereas it is localising if the weight is the indicator function of a one-sided interval, e.g. $ w(z) = \one\{z \geq t\} $. Figure \ref{fig:Deformations} helps to appreciate why this is the case. If the weight is one-sided, then the chaining function maps all elements in $ \{ w = 0 \} $ to a single point. On the other hand, if the weight is an indicator of a compact interval, then points in $ \{ w = 0 \} $ will be mapped to one of two values, depending on where they lie in relation to the interval. As a result, the twCRPS in this case depends not only on the forecast distribution within the region of interest, but also on the measure assigned to outcomes above and below the interval bounds. Nonetheless, if the weight is an indicator function of a one-sided interval, \citet[][Theorem 4]{Holzmann2017} demonstrate that not only is the twCRPS localising, but it is also strictly locally proper.

\subsection{Weighted scores for ensemble forecasts}

Equations \ref{eq:twcrps2} and \ref{eq:owcrps2} allow for straightforward computation of the weighted scoring rules when forecasts are in the form of an empirical distribution function, or an ensemble. In this case, we assume the predictive distribution is a sum of step functions at the $ M $ ensemble members $ x_{1}, ..., x_{M} \in \R $, i.e.~$ F_{ens}(x) = \sum_{m = 1}^{M} \one\{x_{m} \leq x \} / M$. \cite{Gneiting2007} remark that the kernel score representation of the CRPS makes it ``particularly convenient when $ F $ is represented by a sample, possibly based on Markov chain Monte Carlo output or forecast ensembles," which is ordinarily the case in weather and climate forecasting \citep[e.g.][]{Leutbecher2008}. Using the representation of the twCRPS as a kernel score, it is possible to express the threshold-weighted CRPS for an ensemble forecast in the form 
\begin{equation}\label{eq:twCRPSens}
    {\rm twCRPS}(F_{ens}, y; v) = \frac{1}{M}\sum_{m=1}^{M} |v(x_{m}) - v(y)| - \frac{1}{2M^{2}}\sum_{m=1}^{M}\sum_{j=1}^{M} |v(x_{m}) - v(x_{j})|,
\end{equation}
which can be verified by substituting $ F_{ens} $ into Equation \ref{eq:twcrps}. Substituting the identity for $v$ in Equation \ref{eq:twCRPSens} recovers the corresponding well-known expression for the CRPS as a special case. From an implementation standpoint, this expression has the benefit that it involves only a (typically straightforward) transformation of the ensemble members and observations before applying existing methods and software to calculate the CRPS.

\bigskip

Similarly, Equation \ref{eq:owcrps2} permits the following representation of the outcome-weighted CRPS:
\begin{equation*}
    \mathrm{owCRPS}(F_{ens}, y; w) = \frac{1}{M\bar{w}}\sum_{m=1}^{M}| x_{m} - y| w(x_{m}) w(y)  - \frac{1}{2M^{2}\bar{w}^{2}}\sum_{m=1}^{M}\sum_{j=1}^{M} | x_{m} - x_{j} | w(x_{m}) w(x_{j}) w(y),
\end{equation*}
where $ \bar{w} = \sum_{m=1}^{M} w(x_{m})/M $. Note, however, that this expression can result in an undefined score if the weight function is not strictly positive, since $ \bar{w} $ could equal zero: for example, if the weight $ w(z) = \one\{z \geq t\} $ is employed but all ensemble members fall below the threshold $ t $. This is more generally the case for continuous forecast distributions that assign zero probability to $ \{ w > 0 \} $, meaning $ F_{w} $ is undefined, though this will be more prevalent when dealing with finite ensembles. 

\bigskip

\section{Making new kernel scores from old}
\label{section:MNKSFO}


The previous section places existing weighted versions of the CRPS into the framework of kernel scores. \cite{Rasmussen2006} list several operations under which the positive definiteness of a kernel is conserved, and this desirable property has also been studied in the case of negative definite and c.n.d.~kernels \citep[see e.g.][]{Berg1984, Chiles2009}. For example, the sum of several c.n.d.~kernels is itself c.n.d., while it is straightforward to verify that if $ \rho $ is a c.n.d.~kernel, then $ \rho(v(x), v(x^{\prime})) $, for some $ v: \mathcal{X} \to \mathcal{X} $, is also a c.n.d.~kernel. If a kernel is additionally negative definite, then multiplication by a non-negative deterministic function will also yield a negative definite kernel. This permits the construction of flexible kernel scores that can be employed to assess forecasts in a range of different settings.

\bigskip

The threshold-weighted CRPS discussed in the previous section is one example of this. It can be generalised to construct a class of threshold-weighted kernel scores by replacing the Euclidean distance with an arbitrary c.n.d.~kernel.
\begin{defin}\label{def:twks}
    Let $ \rho $ be a c.n.d.~kernel on $ \mathcal{X} $ and let $ v: \mathcal{X} \to \mathcal{X} $ be a measurable function. We define the \emph{threshold-weighted kernel score} with kernel $ \rho $ and chaining function $ v $ as
    \begin{equation}\label{eq:twks_def}
        \mathrm{tw}S_{\rho}(P, y; v) = \mathbb{E}_{P}[\rho(v(X), v(y))] - \frac{1}{2}\mathbb{E}_{P}[\rho(v(X), v(X^{\prime}))] - \frac{1}{2}\rho(v(y), v(y)),
    \end{equation}
    where $ X, X^{\prime} \sim P \in \mathfrak{M} $ are independent, $y \in \mathcal{X}$, and it is assumed that all expectations are finite.
\end{defin}

\bigskip

Theorem \ref{theorem:Proper} implies that if $\rho$ is c.n.d.~with $ \rho(x, x) = 0 $ for all $ x \in \mathcal{X} $, then the threshold-weighted kernel score with kernel $ \rho $ and chaining function $ v $ is proper with respect to $\mathcal{M}_{\tilde{\rho}} $, where $ \tilde{\rho}(x, x^{\prime}) = \rho(v(x), v(x^{\prime})) $. If $\rho$ is negative definite, then the score is proper with respect to $\mathcal{M}^{\tilde{\rho}} $. If the kernel score $ S_{\rho} $ is strictly proper, then it is also possible to characterise the chaining functions that preserve this strict propriety.

\begin{propo}\label{prop:twks}
Let $ \rho $ be a c.n.d.~kernel on $ \mathcal{X}$ and let $ v: \mathcal{X} \to \mathcal{X} $ be a measurable function. If $ S_{\rho} $ is strictly proper with respect to $ \mathcal{M}_{\rho} $ $ (\mathcal{M}^{\rho}) $, then $\mathrm{tw}S_{\rho}( \cdot, \cdot; v) $ is strictly proper with respect to $ \mathcal{M}_{\tilde{\rho}} $ $ (\mathcal{M}^{\tilde{\rho}}) $ if and only if the chaining function $ v $ is injective.
\end{propo}

It could be argued that the strict propriety of a weighted scoring rule is often not of primary concern, since interest is typically only on a subset of possible outcomes: the set $ \{ w > 0 \} $, given the chosen weight function. However, threshold-weighted kernel scores require the specification of a chaining function, which may or may not be associated with a measure, or weight, and there is no canonical way to derive a chaining function that corresponds directly to a given weight. Nevertheless, if a certain weight function has been chosen, it is possible to characterise the chaining functions for which a threshold-weighted kernel score is localising and strictly locally proper.

\begin{propo}\label{prop:twks_lp}
Let $ \rho $ be a c.n.d. kernel on $ \mathcal{X}$ such that $ \rho(x, x) = 0 $ for all $ x \in \mathcal{X} $, let $ w $ be a weight function, and let $ v: \mathcal{X} \to \mathcal{X} $ be a measurable function. Then, $\mathrm{tw}S_{\rho}(\cdot, \cdot; v) $ is localising with respect to $ w $ if and only if $ \rho(v(z), v(\cdot)) = \rho(v(z^{\prime}), v(\cdot)) $ for all $ z, z^{\prime} \in \{w = 0\} $.
\end{propo}
Hence, whether or not a threshold-weighted kernel score is localising with respect to a given weight will depend on the choice of chaining function. 

\begin{remark}
The requirement $ \rho(v(z), v(\cdot)) = \rho(v(z^{\prime}), v(\cdot)) $ for all $ z, z^{\prime} \in \{w = 0\} $ can easily be satisfied by choosing a chaining function such that $ v(z) = v(z^{\prime}) = x_{0} $ for all $ z, z^{\prime} \in \{w = 0\} $, and some $x_0 \in \mathcal{X}$. If $ \rho $ is strictly c.n.d., then it is straightforward to show that this is implied by the requirement. For such a chaining function, we say that the threshold-weighted kernel score is centred at $ x_{0} $.
\end{remark}

\begin{propo}\label{prop:twks_slp}
Let $ \rho $ be a c.n.d. kernel on $ \mathcal{X}$, let $ w $ be a weight function, and let $ v: \mathcal{X} \to \mathcal{X} $ be a measurable function. If $ S_{\rho} $ is strictly proper with respect to $ \mathcal{M}_{\rho} $ ($ \mathcal{M}^{\rho} $), then $\mathrm{tw}S_{\rho}( \cdot, \cdot; v) $ is strictly locally proper with respect to $ \mathcal{M}_{\tilde{\rho}} $ ($\mathcal{M}^{\tilde{\rho}}$) if and only if it is localising and the restriction of $ v $ to $ \{ w > 0 \} $ is injective.
\end{propo}

In contrast to threshold-weighted kernel scores, outcome-weighted scores can be generalised to arbitrary c.n.d.~kernels whilst maintaining a direct connection to the weight function. In particular, similarly to the outcome-weighted CRPS in Equation \ref{eq:owcrps2}, we define outcome-weighted kernel scores as follows; they are a special case of the weighted scores proposed by \cite{Holzmann2017}. 
\begin{defin}\label{def:owks}
    Let $ \rho $ be a c.n.d.~kernel on $ \mathcal{X} $ and let $ w $ be a weight function. Let $ P \in \mathcal{M}_{\rho} $ such that $ \mathbb{E}_{P}[w(X)] > 0$. We define the \emph{outcome-weighted kernel score} with kernel $ \rho $ and weight function $ w $ as
    \begin{equation}\label{eq:owks}
        \mathrm{ow}S_{\rho}(P, y; w) = \frac{1}{C_{w}(P)}\mathbb{E}_{P}[\rho(X, y)w(X)w(y)] - \frac{1}{2C_{w}(P)^{2}}\mathbb{E}_{P}[\rho(X, X^{\prime})w(X)w(X^{\prime})w(y)] - \frac{1}{2}\rho(y, y)w(y),
    \end{equation}
     where $ X, X^{\prime} \sim P $ are independent, $ y \in \mathcal{X} $, and $ C_{w}(P) = \mathbb{E}_{P}[w(X)] $.
\end{defin}

As mentioned for the owCRPS, outcome-weighted kernel scores provide a means of weighting existing kernel scores, but they themselves do not fit into the kernel score framework. The results of \cite{Holzmann2017} discussed in Section \ref{section:LocalProp} clarify when outcome-weighted kernel scores are proper, localising, proportionally locally proper, and how they can be modified to be strictly locally proper. 

\bigskip

It is well-known that if $ \rho $ is a negative definite kernel and $ w $ is a weight function, then $ \check{\rho}(x, x^{\prime}) = \rho(x, x^{\prime})w(x)w(x^{\prime}) $ is also a negative definite kernel. We therefore propose constructing weighted scoring rules based on this weighted kernel, and we label such scores vertically re-scaled kernel scores.
\begin{defin}\label{def:nsks}
Let $ \rho $ be a c.n.d.~kernel on $ \mathcal{X} $ and let $ w $ be a weight function.
\begin{enumerate}[(i)]
    \item If $ \rho $ is negative definite, we define the \emph{vertically re-scaled kernel score} with kernel $ \rho $ and weight function $ w $ as
    \begin{equation}\label{eq:nsks}
        \mathrm{vr}S_{\rho}(P, y; w) = \mathbb{E}_{P}[\rho(X, y)w(X)w(y)] - \frac{1}{2}\mathbb{E}_{P}[\rho(X, X^{\prime})w(X)w(X^{\prime})] - \frac{1}{2}\rho(y, y)w(y)^{2},
    \end{equation}
    where $ X, X^{\prime} \sim P \in \mathcal{M}^{\rho} $ are independent and $ y \in \mathcal{X} $.
    \item If $ \rho $ satisfies $ \rho(x, x) = 0 $ for all $ x \in \mathcal{X} $, we define the \emph{vertically re-scaled kernel score} with kernel $ \rho $, weight function $ w $, and centre $ x_{0} \in \mathcal{X} $ as 
    \begin{equation}\label{eq:nsks2}
    \begin{split}
        \mathrm{vr}S_{\rho}(P, y; w, x_{0}) = & \mathbb{E}_{P}[\rho(X, y)w(X)w(y)] - \frac{1}{2}\mathbb{E}_{P}[\rho(X, X^{\prime})w(X)w(X^{\prime})] \\
        & + (\mathbb{E}_{P}[ \rho(X, x_{0})w(X) ] - \rho(y, x_{0})w(y) ) (\mathbb{E}_{P}[w(X)] - w(y)),
    \end{split}
    \end{equation}
    where $ X, X^{\prime} \sim P \in \mathcal{M}_{\rho} $ are independent and $ y \in \mathcal{X} $.
\end{enumerate} 
\end{defin}

\bigskip
 
Since $ \check{\rho} $ is itself a negative definite kernel and $ \mathcal{M}^{\rho} \subset \mathcal{M}^{\check{\rho}} $, it follows that the associated kernel score in Equation \ref{eq:nsks} is proper with respect to $ \mathcal{M}^{\rho} $. However, although multiplication by a non-negative function preserves the negative definiteness of a kernel, if $ \rho $ is only c.n.d., then it is not necessarily the case that $ \check{\rho} $ will be. On the other hand, if $ \rho $ is a c.n.d.~kernel with $ \rho(x, x) = 0 $ for all $ x \in \mathcal{X} $, then
\begin{equation}\label{eq:rhostar}
    \rho^{*}(x, x^{\prime}) = \rho(x, x^{\prime}) - \rho(x, x_{0}) - \rho(x^{\prime}, x_{0}) 
\end{equation}
will be negative definite, for arbitrary $ x_{0} \in \mathcal{X} $ \citep[Lemma 2.1]{Berg1984}. Since $ \rho^{*} $ is negative definite, it follows that $ \rho^{*}(x, x^{\prime})w(x)w(x^{\prime}) $ is also negative definite, and this kernel is used in Equation \ref{eq:nsks2} to construct the vertically re-scaled kernel score centred at $ x_{0} $. Proposition 20 in \cite{Sejdinovic2013} can be used to show that this score is proper with respect to $\mathcal{M}_\rho$.

\bigskip

Regardless of whether $ \rho $ is negative definite or not, the unweighted kernel score is recovered by choosing $ w(z) = 1 $, and does not depend on $ x_{0} $. In general, however, the vertically re-scaled kernel score will depend on $ x_{0} $. Although it is not immediately obvious how the choice of $ x_{0} $ will affect the score's behaviour in practice, it does not alter the theoretical properties of this weighted score, and for the applications in Section \ref{section:wES} there is always a canonical choice. If a vertically re-scaled kernel score is strictly proper with respect to a class of distributions that contains Dirac measures, then $\{w=0\}$ can contain at most one element. Ensuring strict propriety of a vertically re-scaled score requires stronger assumptions.

\begin{propo}\label{prop:nsks}
Let $ \rho $ be a negative definite kernel on $ \mathcal{X} $ and let $ w > 0 $ be a weight function. If $-\rho$ is strictly integrally positive definite with respect to the maximal possible set of signed measures in the sense of \citet[Definition 2.1]{Steinwart2019}, then $ \mathrm{vr}S_{\rho}(\cdot, \cdot; w) $ is strictly proper with respect to $ \mathcal{M}^{\rho} $.
\end{propo}

If $-\rho$ is a bounded continuous strictly positive definite function on $\mathbb{R}^d$ for any $d \ge 1$, then the requirement in Proposition \ref{prop:nsks} is satisfied. However, if $\rho$ is a c.n.d.~kernel with $\rho(x,x) = 0$, $x \in \mathcal{X}$, applying Proposition \ref{prop:nsks} to $\rho^*$ at \eqref{eq:rhostar} is not always possible since the literature on kernel embedding with unbounded kernels is limited.

\bigskip

It can be seen immediately from their definition that vertically re-scaled kernel scores depend on the forecast $ P $ only through its restriction to the set $ \{ w > 0 \} $, and they are therefore localising. Slightly generalizing Proposition \ref{prop:nsks}, we obtain the following result. 

\begin{propo}\label{prop:nsks_lp}
Let $ \rho $ be a negative definite kernel on $ \mathcal{X} $ and let $ w $ be a weight function. If $-\rho$ is strictly integrally positive definite with respect to the maximal possible set of signed measures on $\{w > 0\}$ in the sense of \citet[Definition 2.1]{Steinwart2019}, then $ \mathrm{vr}S_{\rho}(\cdot, \cdot; w) $ is strictly locally proper with respect to $ \mathcal{M}^{\rho} $.
\end{propo}

Hence, vertically re-scaled kernel scores provide a direct means of obtaining a strictly locally proper scoring rule. Furthermore, they also fit into the kernel score framework. While threshold-weighted kernel scores deform the inputs of the kernel, vertically re-scaled kernel scores weight the kernel's output. However, for particular weight and chaining functions, vertically re-scaled and threshold-weighted kernel scores are equivalent.

\begin{propo}\label{prop:nsks_twks}
Let $ \rho $ be a c.n.d.~kernel on $ \mathcal{X} $ with $ \rho(x, x) = 0 $ for all $ x \in \mathcal{X} $, let $ w $ be a weight function such that $ w(x) \in \{0, 1\} $ for all $ x \in \mathcal{X} $, and let $ x_{0} \in \mathcal{X} $. Consider the chaining function $ v(x) = xw(x) + x_{0}(1 - w(x))$,  $ x \in \mathcal{X} $. Then, the threshold-weighted kernel score with kernel $ \rho $ and chaining function $ v $ equals the vertically re-scaled kernel score with kernel $ \rho $, weight function $ w $, and centre $ x_{0} $.
\end{propo}

Given a kernel score, this section has described three possible approaches that can be used to weight the scoring rule in order to emphasise particular outcomes. Since these approaches apply to the entire class of kernel scores, they are applicable in a wide range of settings. As an example of this, in the following section, we investigate the application of kernel scores in a multivariate context and use results from this section to introduce weighted versions of popular multivariate scoring rules.

\section{Weighted multivariate scoring rules}
\label{section:wES}


\subsection{Energy and variogram scores}

Let $ \mathcal{X} = \R^{d} $ and let $ \mathcal{M} $ denote the set of Borel probability measures on $ \R^{d} $. Forecast verification in a multivariate setting is significantly less developed than in the univariate case \citep{Gneiting2014}. In particular, there are relatively few recognised scoring rules to quantify the accuracy of multivariate forecasts. The logarithmic score can be used to assess multivariate predictive densities, but multivariate forecasts are commonly in the form of a finite ensemble. Applying the logarithmic score to a multivariate normal density recovers the Dawid-Sebastiani score \citep{Dawid1999}, which evaluates forecasts only through their first two moments. Although this makes the score readily applicable to ensemble forecasts, it can become uninformative when the number of dimensions under consideration is large compared with the number of ensemble members. Hence, two alternative multivariate scoring rules are commonly preferred in practice: the energy score and the variogram score. 

\bigskip

The energy score is generally defined as 
\begin{equation}\label{eq:ESkern}
    {\rm ES}_{\beta}(P, y) = \mathbb{E}_{P}||X - y||^{\beta} - \frac{1}{2}\mathbb{E}_{P}||X - X^{\prime}||^{\beta},
\end{equation} 
where $ || \cdot || $ is the Euclidean norm and the exponent $ \beta \in (0, 2) $ is typically set to one \citep{GneitingRaftery2007}. Here, and throughout this section, we assume that all relevant expectations are finite. Clearly, Equation \ref{eq:ESkern} defines a kernel score associated with the c.n.d.~kernel $ \rho(x, x^{\prime}) = ||x - x^{\prime}||^{\beta} $, for $ x, x^{\prime} \in \R^{d} $, and the energy score thus generalises the CRPS to multiple dimensions. The energy score is strictly proper with respect to $ \{ P \in \mathcal{M} : \mathbb{E}_{P}||X||^{\beta} < \infty \} $.

\bigskip

The results presented in the previous section allow us to generate three distinct weighted energy scores. Firstly, Definition \ref{def:twks} can be used to construct a threshold-weighted energy score, which provides a natural extension of the threshold-weighted CRPS to higher dimensions:
\begin{equation*}
    {\rm twES}_{\beta}(P, y; v) = \mathbb{E}_{P}||v(X) - v(y)||^{\beta} - \frac{1}{2}\mathbb{E}_{P}||v(X) - v(X^{\prime})||^{\beta},
\end{equation*}
where $ v: \R^{d} \to \R^{d} $ is a chaining function. Alternatively, applying Equation \ref{eq:owks} to the energy score recovers the outcome-weighted energy score proposed by \cite{Holzmann2017}:
\begin{equation*}
    {\rm owES}_{\beta}(P, y; w) = \frac{1}{C_{w}(P)}\mathbb{E}_{P}[||X - y||^{\beta}w(X)w(y)] - \frac{1}{2C_{w}(P)^{2}}\mathbb{E}_{P}[||X - X^{\prime}||^{\beta}w(X)w(X^{\prime})w(y)],
\end{equation*}
where $w:\mathbb{R} \to [0,1]$ is a weight function.
Similarly, Definition \ref{def:nsks} can be used to introduce a vertically re-scaled energy score. Since the kernel used in the energy score is only c.n.d., this requires choosing a point $ x_{0} \in \R^{d} $ at which to centre the weighted score. The natural choice is $x_0 = 0$: 
\begin{equation*}
\begin{split}
    {\rm vrES}_{\beta}(P, y; w) = & \mathbb{E}_{P}\left[ ||X - y||^{\beta} w(X)w(y) \right] - \frac{1}{2}\mathbb{E}_{P} \left[ ||X - X^{\prime}||^{\beta} w(X)w(X^{\prime}) \right] \\
    & + (\mathbb{E}_{P} \left[ ||X - x_{0}||^{\beta} w(X) \right] - ||y - x_{0}||^{\beta}  w(y))(\mathbb{E}_{P} \left[ w(X) \right] - w(y)).
\end{split}
\end{equation*}
While the results in Section \ref{section:MNKSFO} provide conditions on the weight and chaining functions that ensure strict (local) propriety of the threshold-weighted and outcome-weighted energy score, we can only guarantee propriety for the vertically re-scaled energy score.
\bigskip

Although the energy score is possibly the most widely implemented multivariate scoring rule, several studies have found evidence to suggest that it is over-sensitive to marginal forecast performance \citep[e.g.][]{Pinson2013}. \cite{Scheuerer2015} argue that since marginal performance can be assessed using univariate scoring rules, multivariate assessment should instead focus on evaluating the forecast's dependence structure. To this end, the authors introduce the variogram score as an alternative multivariate scoring rule. Given an observation $ y = (y_{1}, \dots, y_{d}) \in \R^{d} $, the variogram score of order $ p > 0 $ is defined as
\begin{equation}\label{eq:VS}
    {\rm VS}_{p}(P, y) = \sum_{i=1}^{d}\sum_{j=1}^{d} h_{i, j} (\mathbb{E}_{P}|X_{i} - X_{j}|^{p} - |y_{i} - y_{j}|^{p})^{2},
\end{equation}
where $ X = (X_{1}, \dots, X_{d}) \sim P$, and $ h_{i, j} \in [0, 1] $ are non-negative scaling parameters. In contrast to the energy score, the variogram score is proper with respect to the set $ \{ P \in \mathcal{M} : \mathbb{E}_{P}|X_{i}|^{2p} < \infty \hspace{0.1cm} \mathrm{for} \hspace{0.1cm} \mathrm{each} \hspace{0.1cm} i = 1, \dots, d \} $, but is not strictly proper \citep{Scheuerer2015}. Nonetheless, it is straightforward to verify that the variogram score is also a kernel score, corresponding to the c.n.d.~kernel
\begin{equation*}
    \rho(x, x^{\prime}) = \sum_{i=1}^{d} \sum_{j=1}^{d} h_{i,j} (|x_{i} - x_{j}|^{p} - |x_{i}^{\prime} - x_{j}^{\prime}|^{p})^{2}, 
\end{equation*} 
where $ x = (x_{1}, \dots, x_{d}), x^{\prime} = (x_{1}^{\prime}, \dots, x_{d}^{\prime}) \in \R^{d} $. The variogram score can thus also be expressed as 
\begin{equation*}
    {\rm VS}_{p}(P, y) = \mathbb{E}_{P} \left[ \sum_{i=1}^{d}\sum_{j=1}^{d} h_{i, j} (|X_{i} - X_{j}|^{p} - |y_{i} - y_{j}|^{p})^{2} \right] - \frac{1}{2} \mathbb{E}_{P} \left[ \sum_{i=1}^{d}\sum_{j=1}^{d} h_{i, j} (|X_{i} - X_{j}|^{p} - |X^{\prime}_{i} - X^{\prime}_{j}|^{p})^{2} \right],
\end{equation*}
where $ X, X^{\prime} \sim P $ are independent.

\bigskip

The variogram score was introduced as a multivariate scoring rule that is more sensitive to errors in the forecast's dependence structure than the energy score, and hence is itself an example of how the kernel within the kernel score framework can be chosen in order to emphasise particular features of the forecasts. In addition, just as we have introduced weighted versions of the energy score, threshold-weighted, outcome-weighted, and vertically re-scaled versions of the variogram score can also easily be introduced. For example, the threshold-weighted variogram score of order $ p $ with chaining function $ v $ is
\begin{equation*}
    {\rm twVS}_{p}(P, y; v) = \sum_{i=1}^{d}\sum_{j=1}^{d} h_{i, j} (\mathbb{E}_{P}|v(X)_{i} - v(X)_{j}|^{p} - |v(y)_{i} - v(y)_{j}|^{p})^{2}.
\end{equation*}
Note, however, that since the variogram score is not strictly proper, the outcome-weighted version of this score is localising and proper, but not necessarily proportionally locally proper.

\bigskip

The energy score and variogram score are established kernel scores. The kernel score framework also permits the introduction of novel kernel scores by choosing an appropriate kernel. We illustrate this here by introducing a scoring rule based on the inverse multiquadric kernel \citep{Micchelli1984, Scholkopf2002}. In particular, we define the inverse multiquadric score (IMS) as
\begin{equation*}
    \mathrm{IMS}(P, y) = \mathbb{E}_{P} \left[ -(1 + ||X - y||^{2})^{-\frac{1}{2}} \right] - \frac{1}{2} \mathbb{E}_{P} \left[ -(1 + ||X - X^{\prime}||^{2})^{-\frac{1}{2}} \right] + \frac{1}{2},
\end{equation*} 
where $ X, X^{\prime} \sim P $ are independent.

\bigskip

The IMS can be used to assess both univariate and multivariate forecasts, and weighted versions of this score can be constructed using the results presented in the previous section. The strict propriety of the IMS with respect to the entire class $\mathcal{M}$ follows from the fact that the inverse multiquadric kernel is strictly positive definite and bounded. Boundedness of the kernel is in contrast to the kernels used within the CRPS, energy score, and variogram score. In particular, it implies strict local propriety of the vertically re-scaled IMS. By comparing the IMS to these established scores, we can see to what extent these properties of the kernel affect the behaviour of the resulting kernel score.

\subsection{Weight and chaining functions}

Thus far, we have argued that applying a weighted scoring rule is often synonymous with choosing a suitable kernel to employ within the kernel score framework. A natural question then arises regarding what weight or chaining function, and hence what kernel, to choose for a given problem. In this section, we consider the case where the outcome space is the $ d $-dimensional Euclidean space, $ \R^{d} $, and discuss possible weight and chaining functions that could be used within the weighted multivariate scores introduced above in order to evaluate forecasts made for high-impact events.

\bigskip

\begin{figure}
  \begin{subfigure}{0.31\textwidth}
    \includegraphics[width=\linewidth]{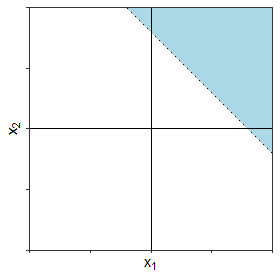}
    \caption{$ w(z) = \one\{z_{1} + z_{2} \geq t \} $}
    \label{fig:MVweights_a}
  \end{subfigure}
  \hspace*{\fill} 
    \begin{subfigure}{0.31\textwidth}
    \includegraphics[width=\linewidth]{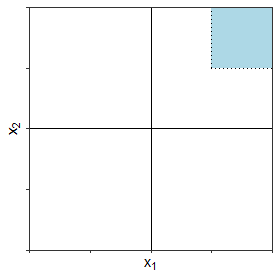}
    \caption{$ w(z) = \one\{z_{1} \geq t_{1}, z_{2} \geq t_{2} \} $}
    \label{fig:MVweights_b}
  \end{subfigure}
  \hspace*{\fill}
  \begin{subfigure}{0.31\textwidth}
    \includegraphics[width=\linewidth]{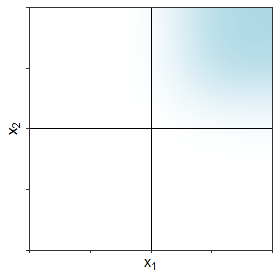}
    \caption{$ w(z) = \boldsymbol{\Phi}(z; \mu, \Sigma) $}
    \label{fig:MVweights_c}
  \end{subfigure}
  \caption{Possible weight functions that could be used to assess forecasts with emphasis on high-impact events in a bivariate setting. A darker shade of blue reflects a higher weight. $ \boldsymbol{\Phi}(z; \mu, \Sigma) $ denotes the cumulative distribution function of the bivariate Gaussian distribution with mean vector $ \mu $ and covariance matrix $ \Sigma $.} 
  \label{fig:MVweights}
\end{figure}

Firstly, consider possible weight functions. In the univariate setting, it is common to assess forecasts with a weight function that only emphasises values above (or below) a chosen threshold, e.g. $ w(z) = \one\{z \geq t\} $. In the multivariate case, this can be extended seamlessly by considering weight functions that are one when a combination of the values along the different dimensions exceeds a threshold, and zero otherwise: $ w(z) = \one\{\sum_{i=1}^{d} b_{i} z_{i} \geq t\} $ for constants $ b_{1}, \dots, b_{d} \in \R $ and a threshold $ t \in \R $. An example of this weight function is displayed in Figure \ref{fig:MVweights_a}. 

\bigskip

However, it may be the case that high-impact events arise from the interaction of several moderate events. For example, moderate rainfall over consecutive days is likely to result in flooding, despite the rainfall on each day not being extreme from a statistical perspective. Hence, one could also consider using a weight function that depends on whether a threshold is exceeded in every dimension, e.g.~$ w(z) = \one\{z_{1} \geq t_{1}, \dots, z_{d} \geq t_{d}\} $ with $ t_{1}, \dots, t_{d} \in \R $. Such weight functions can be interpreted as indicator functions of orthants in Euclidean space, as illustrated for the bivariate case in Figure \ref{fig:MVweights_b}.

\bigskip

Both of these weights are based on indicator functions, meaning they are not strictly positive. As such, the outcome-weighted and vertically re-scaled scoring rules constructed using these weights will not be strictly proper. The final column of Figure \ref{fig:Deformations} provides an example of a univariate weight function that is strictly positive, the Gaussian distribution function. This can also readily be extended to higher dimensions by using a weight function equal to a continuous multivariate distribution function. In this case, the weight changes smoothly over $ \R^{d} $, whilst still allowing emphasis to be placed on certain outcomes, as illustrated in Figure \ref{fig:MVweights_c}.

\bigskip

If a threshold-weighted kernel score is to be implemented, it is necessary to specify a chaining function, rather than a weight, which is a less trivial task. In the univariate case, a weight function can typically be converted to a suitable chaining function due to the integral representation of the threshold-weighted CRPS (Remark \ref{remark:twcrps}). More generally, however, there is no canonical way to derive a chaining function given a particular weight. 

\bigskip

Proposition \ref{prop:twks_lp} states that the threshold-weighted kernel score will be localising if the chaining function maps all points in $ \{ w = 0 \} $ to a single value $ x_{0} \in \R^{d} $, while Proposition \ref{prop:twks_slp} states that if the chaining function is additionally injective on $ \{ w > 0\} $, then the score will be strictly locally proper. If all points in $ \{ w > 0 \} $ receive the same weight, as is the case for indicator-based weight functions, then one option is to employ the function
\begin{equation}\label{eq:tw_loc_deform}
v(z) = 
\begin{cases}
  z \text{ if } z \in \{ w > 0\}, \\
  x_{0} \text{ if } z \in \{ w = 0\} ,
 \end{cases}
\end{equation}
for some $x_0 \in \mathbb{R}^d$.
Such a chaining function maintains correspondence with the twCRPS when the weight is a one-sided interval: for example, when $ d = 1 $ and $ w(z) = \one\{z \geq t\} $, choosing $ x_{0} = t $ recovers the chaining function $ v(z) = {\rm max}(z, t) $, as presented in Figure \ref{fig:Deformations}. 

\bigskip

On the other hand, if the weight is not constant on $ \{ w > 0 \} $, then what chaining function to choose will depend strongly on what the weighting is designed to achieve.  Since we are interested here in high-impact events, consider the case where the weight function is increasing along each dimension, a multivariate distribution function, for example. In the univariate case, if the weight is increasing, then the resulting chaining function is convex. One way to translate this to the multivariate setting is to use a chaining function that is convex along every dimension. 

\bigskip

Following Remark \ref{remark:twcrps}, a possible chaining function that satisfies this is the integral of the weight function along each margin separately, conditional on the values along the other dimensions. For example, the final weight function considered in Figure \ref{fig:MVweights} employs a multivariate Gaussian distribution function. Given a mean vector $ \mu = (\mu_{1}, \dots, \mu_{d}) $ and a diagonal covariance matrix $ \Sigma $ with variances $ \sigma_{1}^{2}, \dots, \sigma_{d}^{2} $, integrating the conditional Gaussian distribution along each dimension yields a chaining function of the form
$$ v(z) = \left( (z_{1} - \mu_{1})\Phi\left(\frac{z_{1} - \mu_{1}}{\sigma_{1}}\right) + \sigma_{1}\phi\left(\frac{z_{1} - \mu_{1}}{\sigma_{1}}\right), \dots, (z_{d} - \mu_{d})\Phi\left(\frac{z_{d} - \mu_{d}}{\sigma_{d}}\right) + \sigma_{d}\phi\left(\frac{z_{d} - \mu_{d}}{\sigma_{d}}\right) \right), $$
which is essentially a component-wise extension of the chaining function presented in the final column of Figure \ref{fig:Deformations}. Moreover, since this chaining function is injective on $ \{w > 0\} $, Proposition \ref{prop:twks_slp} states that the resulting threshold-weighted kernel score will be strictly locally proper.

\bigskip

These are only a few examples of possible weight and chaining functions that could be employed when evaluating forecasts and outcomes on $ \R^{d} $ whilst emphasising high-impact events. Different choices of either function will generate a scoring rule that assesses different aspects of the forecast performance, and, in general, it is a task for a domain expert to choose the appropriate weight or chaining function in order to extract the relevant information for the problem at hand. In the remainder of this section, we examine how the weights and deformations presented above can be used within weighted multivariate scoring rules to assess forecasts made for high-impact events.

\subsection{Simulation study}

\subsubsection{Outline}

In order to understand the properties of the various weighted multivariate scoring rules, we apply them to simulated forecasts and observations. The simulation study is organised as follows. Firstly, a distribution $ G $ is chosen from which to draw 100 independent observations. Secondly, two forecast distributions, $ F_{1} $ and $ F_{2} $, are specified, both of which are linear combinations of the true distribution $ G $, and a mis-specified distribution $ H $:
\begin{equation*}
\begin{split}
    F_{1}(z) = a(z)G(z) + (1 - a(z))H(z), \\
    F_{2}(z) = (1 - a(z))G(z) + a(z)H(z),
\end{split}
\end{equation*}
where $ F_{1}, F_{2}, G $ and $ H $ all denote distributions on $ \R^{d} $, while $ a: \R^{d} \to [0, 1] $ is a mixing function. 

\bigskip 

\begin{figure}
    \centering
    \includegraphics[width = 0.48\linewidth]{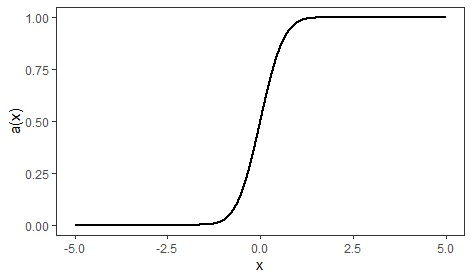}
    \includegraphics[width = 0.48\linewidth]{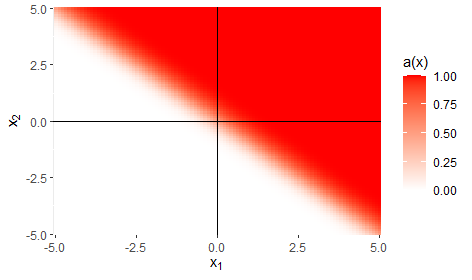}
    \caption{The mixing function $ a(x) $ in the univariate case (Left) and the bivariate case (Right).}
    \label{fig:1D_Weights}
\end{figure}

In the following, $ G $ denotes a standard multivariate Gaussian distribution, while $ H $ is a multivariate Student's $ t $ distribution with four degrees of freedom \citep{Hothorn2001}. As in \cite{Holzmann2017}, the mixing function $ a $ is a univariate Gaussian distribution function with zero mean and standard deviation equal to one half. In the multivariate case, this univariate distribution function is evaluated at $ \sum_{i=1}^{d} z_{i} $. This mixing function is displayed for the univariate and bivariate cases in Figure \ref{fig:1D_Weights}. 


\bigskip

In order to assess the competing forecasts, 100 ensemble members are sampled at random from $ F_{1} $ and $ F_{2} $, and both forecasts are then evaluated at each of the 100 observations via ensemble forecast representations of the various weighted scoring rules. A Diebold-Mariano test \citep{Diebold1995} is applied to the sample mean scores of both forecasts to assess whether or not one forecast outperforms the other when assessed using each score. This process is repeated 1000 times and the proportion of instances that the null hypothesis of equal predictive performance is rejected in favour of each forecast distribution is recorded. The rejection rates for the various scoring rules can then be examined to understand the discriminative behaviour of the different scores. Such a framework has been considered in several previous studies on weighted scoring rules \citep{Diks2011, Lerch2017, Holzmann2017}.

\bigskip

In the univariate case, the forecasts are assessed using the CRPS and the IMS, while in the multivariate case, results are presented for the energy score, the variogram score and the IMS. Threshold-weighted, outcome-weighted, and vertically re-scaled versions of these kernel scores are all considered, as well as outcome-weighted scores that have been complemented with the Brier score in order to make them strictly locally proper, as discussed in Section \ref{section:LocalProp} (Equation \ref{eq:owComp}). 

\bigskip

For simplicity, we consider only indicator-based weight functions within these weighted scores, which are commonly applied in practice. Results are presented for the univariate weight function $ w(z) = \one\{ z \geq t \} $, where interest is on values that exceed a chosen threshold, and for two multivariate, indicator-based weight functions: $ w(z) = \one\{\sum_{i=1}^{d} z_{i} \geq t\} $  and $ w(z) = \one\{ z_{1} \geq t, \dots, z_{d} \geq t \} $. In all cases, we study the results as the threshold $ t $ is changed. 

\bigskip

Equation \ref{eq:tw_loc_deform} is used to construct a chaining function that generates localising threshold-weighted kernel scores. In the first scenario, when $ w(z) = \one\{ z_{1} \geq t, \dots, z_{d} \geq t \} $, the threshold-weighted scores are centred at $ x_{0} = (t, \dots, t) $, while in the second case, when $ w(z) = \one\{\sum_{i=1}^{d} z_{i} \geq t\} $, the scores are centred at $ x_{0} = (t/d, \dots, t/d) $. Where relevant, the vertically re-scaled kernel scores are centred at $ (0, \dots, 0) $. If the kernel in the kernel is only c.n.d., this is equivalent to using a threshold-weighted kernel score centred at the origin (Proposition \ref{prop:nsks_twks}). Hence, comparing the performance of the threshold-weighted and vertically re-scaled scores allows us to assess how the scores depend on the point at which they are centred. For the threshold-weighted variogram score, however, the kernel is insensitive to whether the score is centred at $ (t, \dots, t) $ or $ (0, \dots, 0) $, and hence the results for the localising threshold-weighted variogram score are equivalent to those for the vertically re-scaled variogram score.

\bigskip

For the sake of comparison, these scores are compared to threshold-weighted kernel scores that are non-localising. In this case, the chaining function is chosen to resemble the chaining function in the univariate case. Firstly, consider the weight function $ w(z) = \one\{z_{1} \geq t, \dots z_{d} \geq t \} $. One possible chaining function corresponding to this weight would be to take the component-wise maximum between the point and the threshold:
\begin{equation*}
    v(z) = (\mathrm{max}(z_{1}, t), \dots , \mathrm{max}(z_{d}, t)).
\end{equation*}
As has been discussed, the weight function in this case can be thought of as an orthant in Euclidean space, and this chaining function simply projects any point not contained in this orthant onto the perimeter of the orthant, while leaving the remaining points unchanged. That is, points that lie in the region of interest, $ \{ w > 0 \} $, are unchanged, whereas points outside this region are projected onto the closest point for which the weight equals one.

\bigskip

This approach could similarly be used when considering the weight function $ w(z) = \one\{\sum_{i=1}^{d} z_{i} \geq t\} $, in which case the chaining function becomes
\begin{equation*}
    v(z) = (\mathrm{max}(z_{1}, z_{1} + l), \dots, \mathrm{max}(z_{d}, z_{d} + l)),
\end{equation*}
where $ l = (t - \sum_{i=1}^{d} z_{i})/d  $. Here, any point in $ \{ w = 0 \} $ is moved perpendicular to the plane defined by the weight function, until it reaches a point on the plane.

\subsubsection{Results}

Firstly, consider the univariate case. The forecast $ F_{1} $ is correctly specified in the upper tail, but exhibits a heavy lower tail, whereas the opposite is true for $ F_{2} $. As such, since evaluation is focused on the upper tail of the forecast distributions, for large values of $ t $ the weighted scoring rules should reject the null hypothesis of equal predictive performance in favour of $ F_{1} $. 
\bigskip

\begin{figure}[!t]
    \centering
    \includegraphics[width = 0.48\linewidth]{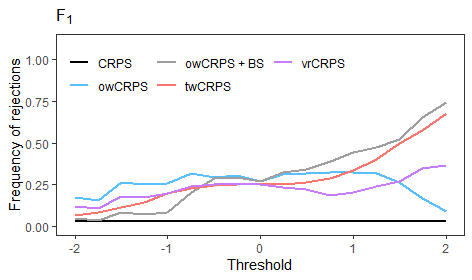}
    \includegraphics[width = 0.48\linewidth]{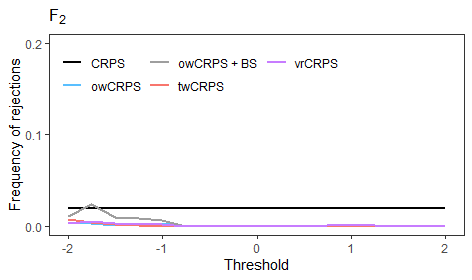}
    \includegraphics[width = 0.48\linewidth]{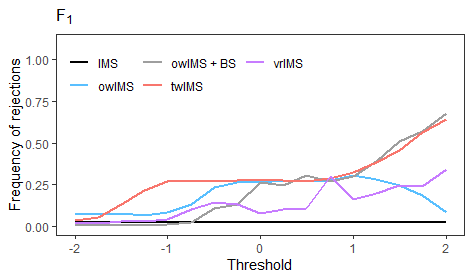}
    \includegraphics[width = 0.48\linewidth]{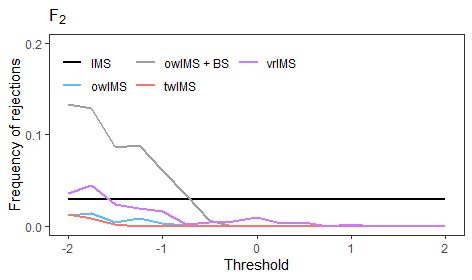}
    \caption{The proportion of instances that a Diebold-Mariano test for equal predictive performance is rejected in favour of $ F_{1} $ (Left) and $ F_{2} $ (Right) for each version of the CRPS (Top) and the IMS (Bottom). The rejection rate is displayed as a function of the threshold used in the weight function $ w(z) = \one\{z \geq t\} $ when evaluating the forecasts. Note the different scales when considering $ F_{1} $ and $ F_{2} $.}
    \label{fig:1D_RejProp}
\end{figure}

While the rejection rate of the unweighted scores is close to 0.025 for all thresholds, Figure \ref{fig:1D_RejProp} illustrates that the frequency of rejections in favour of $ F_{1} $ generally increases with the threshold when a weighted scoring rule is used to assess the forecasts. The outcome-weighted scores perform poorly for larger thresholds, since they are sensitive to the number of observations that exceed the threshold, a result also observed in \cite{Holzmann2017}. Complementing the owCRPS and owIMS with the Brier score generates scores that can better distinguish between the two forecasts. The threshold-weighted scores also perform well in this respect. Comparing the threshold-weighted CRPS to the vertically re-scaled CRPS illustrates the sensitivity of the scores to the point at which the two weighted scores are centred: in this case, centering the scores at the threshold appears more beneficial than centering them at the origin.

\bigskip

The right-hand panel of Figure \ref{fig:1D_RejProp} shows the proportion of rejections in favour of $ F_{2} $. As would be expected, and in contrast to $ F_{1} $, this rejection frequency is close to zero for large thresholds, and increases towards 0.025 as the threshold becomes smaller, mirroring the results presented in \cite{Holzmann2017}. The owIMS complemented with the Brier score, on the other hand, results in a large rejection frequency when the threshold is low.

\bigskip

Consider now results for the bivariate setting. In this case, $ F_{1} $ is close to the true data generating process $ G $ in the upper right quadrant (when the mixing function is close to one), whereas $ F_{2} $ is more similar to the true distribution in the lower left quadrant, meaning the weighted scores should again reject the hypothesis of equal predictive performance in favour of $ F_{1} $ when the threshold $ t $ is large. 

\bigskip

\begin{figure}[t!]
    \centering
        \includegraphics[width = 0.48\linewidth]{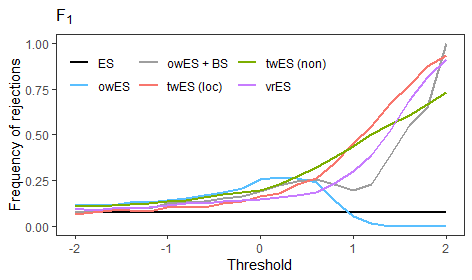}
        \includegraphics[width = 0.48\linewidth]{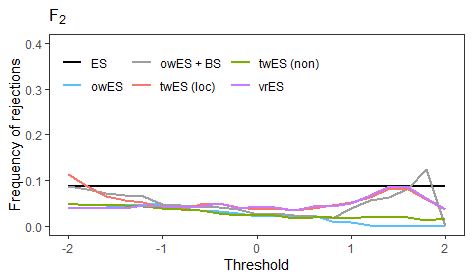}
        \includegraphics[width = 0.48\linewidth]{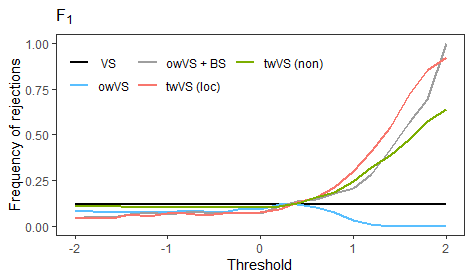}
        \includegraphics[width = 0.48\linewidth]{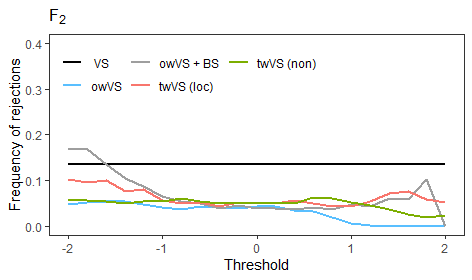}
        \includegraphics[width = 0.48\linewidth]{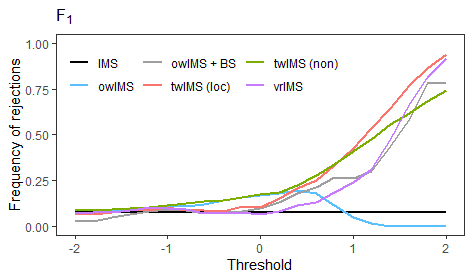}
        \includegraphics[width = 0.48\linewidth]{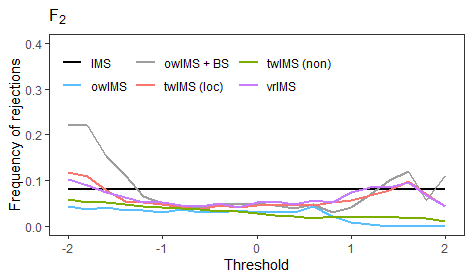}
    \caption{The proportion of instances that a Diebold-Mariano test for equal predictive performance is rejected in favour of $ F_{1} $ (Left) and $ F_{2} $ (Right) for each version of the energy score (Top), variogram score (Middle), and IMS (Bottom). The rejection rate is displayed as a function of the threshold used in the weight function $ w(z) = \one\{z_{1} \geq t, z_{2} \geq t\} $ when evaluating the forecasts. Localising versions of the threshold-weighted scores are denoted by (loc), while non-localising variants are labelled (non). Note the different scales when considering $ F_{1} $ and $ F_{2} $.}
    \label{fig:2D_RejProp_rec}
\end{figure}

Figure \ref{fig:2D_RejProp_rec} displays the rejection frequency in favour of $ F_{1} $ and $ F_{2} $ corresponding to each scoring rule for the weight function $ w(z) = \one\{z_{1} \geq t, z_{2} \geq t\} $. Similar results are observed to those presented in the univariate case. In particular, the energy score, variogram score and the inverse multiquadric score all cannot distinguish between the two forecasts, whereas the weighted scores do, particularly when interest is on relatively large thresholds. The rejection rates corresponding to the outcome-weighted scores increase slightly with the threshold, but then tend towards zero as the number of observations that exceed the threshold decreases. Complementing these scores with the Brier score again generates scoring rules that are capable of identifying the differences between the forecasts. 

\bigskip

The threshold-weighted scores are also adept at capturing the differences in forecast behaviour as the threshold is increased. This is true for both the localising and non-localising variants, though the localising score is generally more discriminative for large thresholds. For the energy score, the vertically re-scaled score is again slightly less informative than the localising threshold-weighted score, suggesting it is preferable to centre these scores close to the threshold of interest.

\bigskip

\begin{figure}[t!]
    \centering
        \includegraphics[width = 0.48\linewidth]{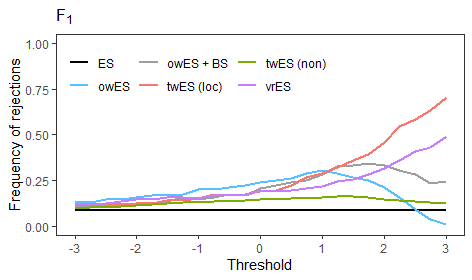}
        \includegraphics[width = 0.48\linewidth]{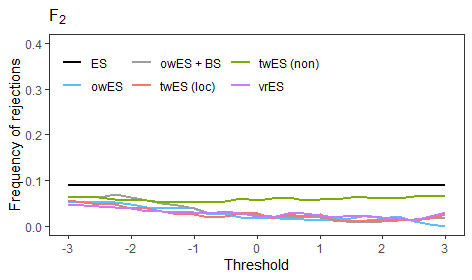}
        \includegraphics[width = 0.48\linewidth]{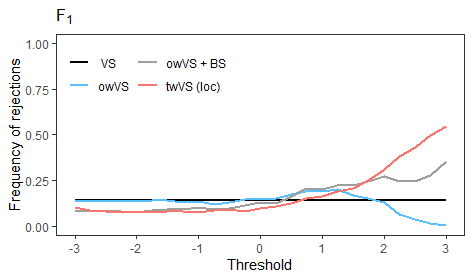}
        \includegraphics[width = 0.48\linewidth]{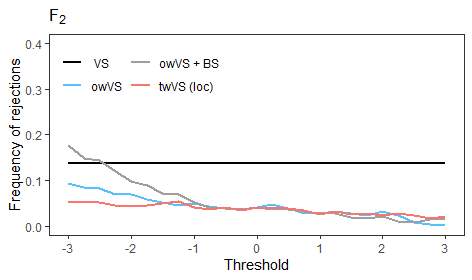}
        \includegraphics[width = 0.48\linewidth]{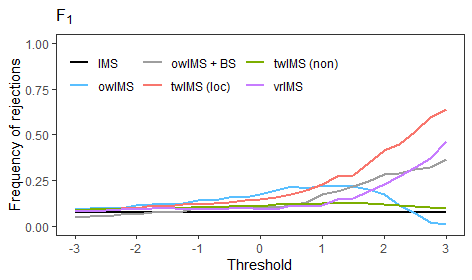}
        \includegraphics[width = 0.48\linewidth]{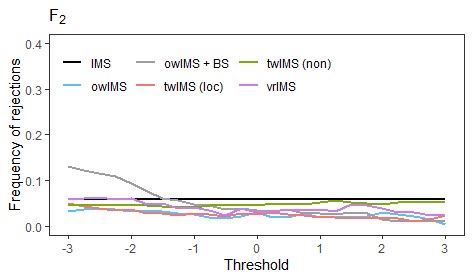}
    \caption{Same as Figure \ref{fig:2D_RejProp_rec} for the weight function $ w(z) = \one\{z_{1} + z_{2} \geq t\} $.}
    \label{fig:2D_RejProp_L1}
\end{figure}

Figure \ref{fig:2D_RejProp_L1} displays the corresponding results for the weight function $ w(z) = \one\{z_{1} + z_{2} \geq t\} $. The weighted scoring rules in this case appear to be less discriminative than in the previous setting, though the conclusions are largely similar. The unweighted multivariate scores cannot distinguish between the two forecasts, whereas the weighted scores do so successfully. The exception to this is the non-localising threshold-weighted scores, which are only marginally more discriminative than the unweighted scores, regardless of the threshold used within the weight function. The reason for this is that the deformation function in this case projects points in $ \{ w = 0\} $ onto the line $ z_{1} + z_{2} = t $, which still contains a lot of information. The resulting score is thus dominated by differences between points in $ \{ w = 0\} $.

\section{Case study}
\label{section:CaseStudy}

\subsection{Introduction}

The simulation study in the previous section demonstrates the utility of weighted multivariate scoring rules when evaluating forecasts made for high-impact events. In this section, we seek to reinforce this by illustrating how these weighted scores can be applied in practice. In particular, the weighted energy, variogram and inverse multiquadric scores described previously are used to evaluate daily rainfall accumulation forecasts across several forecast lead times, with emphasis on events that could lead to flooding. 

\bigskip

Flooding and other associated impacts could manifest, for example, from a large precipitation accumulation on a single day, or from moderate rainfall over consecutive days. Changing the weight used within the weighted multivariate scoring rules allows us to consider these different possibilities when evaluating forecasts. Whether or not an impact occurs will depend not only on the amount of rainfall, but also on other factors, such as the temperature or a location's capabilities to deal with large rainfall accumulations. These external factors are not considered in this study, though the weight or chaining function within the weighted scoring rules could possibly be adjusted to include this information. For example, a weight function could be used that employs different parameters depending on certain covariates or location-specific characteristics.

\bigskip

\begin{figure}
    \centering
    \includegraphics[width = 0.48\linewidth]{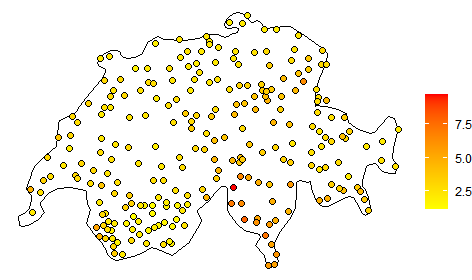}
    \caption{The weather stations across Switzerland and the surrounding area at which precipitation forecasts are considered. The colour of each point reflects the mean daily accumulation at that station, measured in millimetres.}
    \label{fig:PrecMap}
\end{figure}

The daily precipitation accumulation forecasts considered here were issued by the Swiss Federal Office of Meteorology and Climatology's (Meteoswiss) COSMO-E ensemble prediction system. The forecasts and corresponding observations are therefore available at a large number of weather stations across Switzerland. The 245 stations are displayed in Figure \ref{fig:PrecMap}, which also presents the average observed daily accumulation for each station over the period of interest, the four autumns seasons (September to November) between 2016 and 2019. This results in roughly 100,000 pairs of forecasts and observations. The forecasts are initialised at 00 UTC, and their performance is analysed over the three consecutive days following this initialisation time.

\bigskip

\begin{figure}[b!]
    \centering
    \includegraphics[width = 0.48\linewidth]{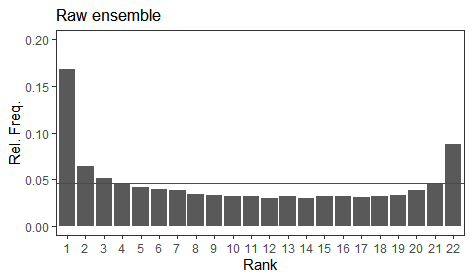}
    \includegraphics[width = 0.48\linewidth]{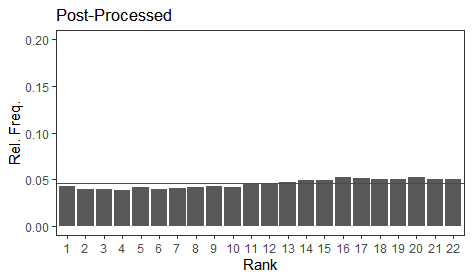}
    \caption{Rank histograms for the COSMO-E prediction system (Left) and the statistical post-processing method (Right) at a lead time of one day. A horizontal red line has been added to indicate perfect calibration. The ranks have been aggregated across all locations and all forecast instances in the test data set, and ties between ranks have been resolved at random.}
    \label{fig:RankHists}
\end{figure}

The COSMO-E prediction system issues forecasts in the form of a 21-member ensemble. However, operational ensemble forecasts made for surface weather variables are commonly found to be overconfident, or under-dispersed, exhibiting less spread than desired. This can be verified using rank histograms, which display the relative frequency that the observed precipitation accumulation is assigned each rank when pooled among the ensemble members. Rank histograms can thus be thought of as an empirical analogue of the probability integral transform (PIT) histogram \citep{Dawid1984, Gneiting2007}. If the ensemble is calibrated, then the observation should be equally likely to assume each possible rank, resulting in a uniform rank histogram. However, the rank histogram for the COSMO-E ensemble forecasts at a lead time of one day (aggregated across all stations) in Figure \ref{fig:RankHists} shows that this is not the case, and the observation tends to fall above or below all ensemble members significantly more often than would be expected from a calibrated forecast. 

\bigskip

As such, it is common for these dynamical forecasts to undergo some form of statistical post-processing. Since the physical mechanisms underlying flooding events may occur on timescales larger than one day, the COSMO-E forecasts are post-processed at individual lead times and then combined using copula approaches to generate multivariate forecasts for the precipitation accumulation for the following three days.

\bigskip

\subsection{Statistical post-processing}

The post-processing approach implemented here follows that proposed by \cite{Scheuerer2015b}, which assumes that the daily precipitation accumulation follows a Gamma distribution that is shifted and censored below at zero; the shifting and censoring yields a zero-inflated distribution that captures the positive probability that the precipitation is exactly zero. The mean and standard deviation of the censored, shifted Gamma distribution are assumed to depend linearly on the ensemble mean and ensemble standard deviation respectively, as is commonly assumed within the Non-homogeneous Regression, or Ensemble Model Output Statistics post-processing framework \citep{Gneiting2005}.

\bigskip

In particular, letting $ Y $ denote the daily precipitation accumulation at a particular station and lead time, and $ x $ the corresponding COSMO-E ensemble forecast, with mean $ \bar{x} $ and standard deviation $ s $, the post-processing model is
\begin{equation*}
    Y | x \sim \Gamma_{0}(\kappa, \theta, \xi), \hspace{1cm} \mu = \kappa \theta = \alpha + \beta \bar{x}, \hspace{1cm} \sigma = \theta \sqrt{\kappa} = \gamma + \delta s, 
\end{equation*}
where $ \Gamma_{0}(\kappa, \theta, \xi) $ denotes the Gamma distribution with shape $ \kappa $ and scale $ \theta $, shifted negatively by $ \xi $ and censored below at zero. The mean of the Gamma distribution is represented by $ \mu $, and the standard deviation by $ \sigma $. These are linked to regression parameters $ \alpha, \beta, \gamma $ and $ \delta $, which, along with $ \xi $, are estimated using maximum likelihood estimation over a training data set. All parameters are constrained to be non-negative using a square-root link function.

\bigskip

The training data set consists of all forecasts issued during the autumn seasons of 2016 and 2017, while the resulting forecasts are then evaluated out-of-sample on the data available during 2018 and 2019. A separate post-processing model is fit to forecasts at each lead time and each station under consideration: the distributional assumptions are the same in each case, but separate sets of model coefficients are estimated. The right-hand panel of Figure \ref{fig:RankHists} demonstrates that this post-processing method successfully re-calibrates the COSMO-E ensemble forecasts for daily precipitation accumulation on average. The calibration of the statistically post-processed forecasts does not change depending on the lead time, whereas the COSMO-E output becomes gradually less under-dispersed as forecast lead time increases.

\bigskip

Flooding could occur due to rainfall events that persist over several days, and in order to anticipate such an event, the forecasts should capture the temporal dependence in the precipitation observations. Hence, the post-processed forecasts for the daily precipitation accumulation at each lead time are combined into a single multivariate forecast distribution. This is achieved using a copula to model the dependence between the precipitation accumulation on consecutive days. Four different copulas are considered.

\bigskip

An independence copula assumes that there is no dependence between the precipitation accumulation on successive days (conditional on the ensemble output), whereas a comonotonic copula conversely assumes perfect dependence, so that heavy rainfall on one day is followed by heavy rainfall on the next day. These copulas therefore serve as convenient baseline approaches to which alternative methods can be compared. The third approach that we consider utilises an empirical copula based on the dependence structure observed in the COSMO-E ensemble forecast; it thus assumes that the numerical weather model correctly simulates the dependence structure observed in reality. This approach, called ensemble copula coupling (ECC), is well-established in the post-processing literature, and is commonly implemented in operational post-processing suites \citep{Schefzik2013}. Finally, we employ a Gaussian copula, which was found here to outperform alternative parametric copula families.

\bigskip

In all cases, 21 ensemble members are generated by sampling from the univariate post-processed distributions at equidistant quantiles, which are then reordered according to the four copulas. Hence, like the COSMO-E output, the resulting forecasts are in the form of a three-dimensional, 21 member ensemble forecast. This implementation of a Gaussian copula differs from previous applications in a post-processing context that draw a random sample from the copula \citep[e.g.][]{Moller2013, Lerch2020}, which is then transformed using the quantile function of the univariate post-processed distributions. The new approach implemented and advocated here, described in detail in the Appendix, ensures that all multivariate post-processing methods exhibit the same marginal forecast performance.

\bigskip

The marginal forecast performance at each lead time is evaluated using the CRPS and the univariate IMS, while the multivariate forecast distributions are assessed with the energy score, the variogram score and the IMS. Since interest is predominantly on forecasts made for events that could lead to flooding, threshold-weighted and vertically re-scaled versions of these kernel scores are also employed, both in a univariate context for the individual daily accumulations, and in a multivariate context for the combined daily accumulations over three days. 

\bigskip

\begin{figure}[t!]
    \centering
    \includegraphics[width = 0.48\linewidth]{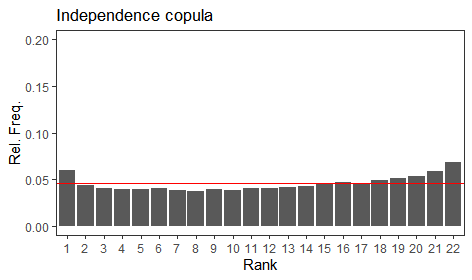}
    \includegraphics[width = 0.48\linewidth]{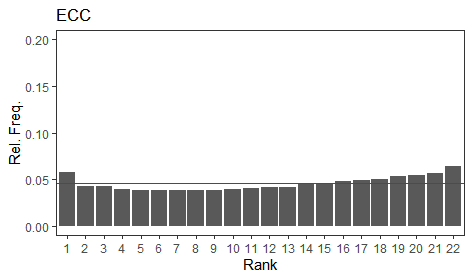}
    \includegraphics[width = 0.48\linewidth]{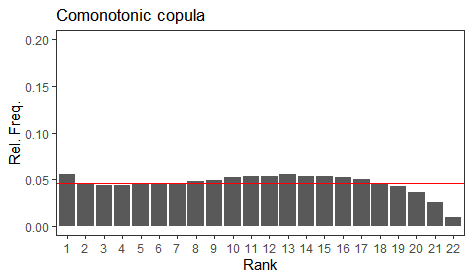}
    \includegraphics[width = 0.48\linewidth]{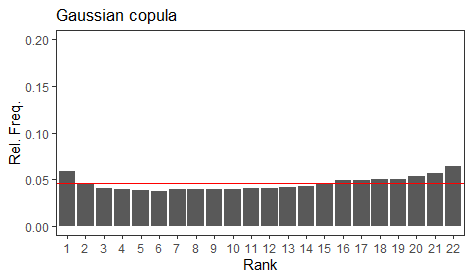}
    \caption{Multivariate rank histograms for the COSMO-E prediction system and the three copula-based multivariate post-processing methods. A horizontal red line has been added to indicate perfect calibration. The ranks have been aggregated across all locations and all forecast instances in the test data set, and ties between ranks have been resolved at random.}
    \label{fig:mvRankHists}
\end{figure}

Before evaluating the accuracy of the multivariate forecasts using these weighted scores, the calibration of the forecasts is assessed using multivariate rank histograms. Multivariate rank histograms provide a natural extension of the rank histograms in Figure \ref{fig:RankHists}, and a uniform histogram is again indicative of a calibrated forecast. Several approaches have been proposed to construct multivariate rank histograms and we implement the approach introduced by \cite{Gneiting2008}, which is presented in Figure \ref{fig:mvRankHists} for the four post-processing approaches. 

\bigskip

In this case, the observation assumes the highest rank if it is larger than the forecast ensemble in all dimensions. Figure \ref{fig:mvRankHists} therefore suggests that the comonotonic copula overestimates the dependence between precipitation on successive days, as expected, while the independence copula, ECC, and Gaussian copula approaches all result in forecasts that slightly underestimate this dependence. Indeed, this is known to be a disadvantage of ECC when forecasting precipitation, since several ensemble members often predict zero precipitation and are then reordered at random \citep{Scheuerer2018}. For further details regarding the interpretation of these multivariate histograms, readers are diverted to \cite{Gneiting2008,ZiegelGneiting2014}. 

\subsection{Weighted scoring rules}

Since the multivariate post-processed forecasts differ only in their choice of copula, all exhibit the same univariate forecast performance. Table \ref{tab:uvScores} displays the CRPS and the IMS for these post-processed forecasts, as well as the COSMO-E model output. Both scores suggest that statistical post-processing is beneficial at all lead times considered here, though the improvement decreases as lead time increases. Table \ref{tab:uvScores} also presents the threshold-weighted and vertically re-scaled versions of these two univariate scores. The weight in this case is an indicator function that emphasises daily precipitation accumulations that exceed 50mm, roughly corresponding to the 99th percentile of the observed accumulations in the training data set. The corresponding chaining function used within the threshold-weighted scores is $ v(z) = \mathrm{max}(z, 50) $. Similarly to the unweighted scores, post-processing improves upon the raw ensemble forecast at all lead times, regardless of the weighted score used to assess the forecasts.

\begin{table}
    \centering
    \begin{tabular}{| c | c c | c c | c c |} 
        \hline
          & \multicolumn{2}{| c |}{1 day} & \multicolumn{2}{| c |}{2 days} & \multicolumn{2}{| c |}{3 days} \\
          \cline{2-7}
          & Raw & Post proc. & Raw & Post proc. & Raw & Post proc. \\
          \hline
          CRPS & 1.30 (0.02) & 1.23 (0.02) & 1.38 (0.02) & 1.32 (0.02) & 1.51 (0.02) & 1.47 (0.02) \\
          twCRPS & 0.262 (0.011) & 0.244 (0.010) & 0.285 (0.011) & 0.264 (0.010) & 0.277 (0.010) & 0.269 (0.010) \\
          vrCRPS & 0.642 (0.019) & 0.609 (0.018) & 0.676 (0.020) & 0.635 (0.019) & 0.680 (0.019) & 0.665 (0.020) \\
         \hline
          IMS & 1.41 (0.01) & 1.30 (0.01) & 1.40 (0.01) & 1.33 (0.01) & 1.47 (0.01) & 1.42 (0.01) \\
          twIMS & 0.167 (0.005) & 0.160 (0.005) & 0.168 (0.004) & 0.161 (0.005) & 0.171 (0.004) & 0.166 (0.005) \\
          vrIMS & 0.132 (0.004) & 0.127 (0.004) & 0.131 (0.004) & 0.126 (0.004) & 0.132 (0.004) & 0.128 (0.004) \\
         \hline
    \end{tabular}
    \caption{Unweighted and weighted univariate scores for the raw COSMO-E output and the post-processed forecasts at each lead time. The scores have been averaged across all locations and all forecast instances in the test data set. Standard errors for the scores are shown in brackets, and the inverse multiquadric scores have been scaled by 10 to ease interpretation.}
    \label{tab:uvScores}
\end{table}

\bigskip

The multivariate forecasts made for the precipitation accumulation across the three lead times are evaluated using the energy score, the variogram score and the IMS, which are presented in Table \ref{tab:mvScores}. The comonotonic copula approach performs worst with respect to all unweighted scores, whereas the COSMO-E ensemble outperforms the post-processed forecasts when evaluated using the variogram score. Since the variogram score is more sensitive to the forecast dependence structure than the energy score and the IMS, this result suggests that any improvements gained by post-processing are principally due to an improved univariate performance. As in Figure \ref{fig:mvRankHists}, the independence copula, ECC, and Gaussian copula approaches all perform similarly, suggesting the COSMO-E output already captures the majority of the dependence between the precipitation accumulation on successive days. 

\bigskip

Results are also presented in Table \ref{tab:mvScores} when evaluating forecasts using threshold-weighted and vertically re-scaled versions of these multivariate scores. Firstly, consider a weight function that is equal to one when the daily precipitation exceeds 25mm on the three consecutive days - this is labelled a ``successive exceedance" in Table \ref{tab:mvScores}. As in the simulation study, two different chaining functions are used within the threshold-weighted energy and variogram scores, one of which results in a localising weighted score, while the other does not. 

\bigskip

The conclusions drawn from the two chaining functions are not the same. The raw ensemble output and the comonotonic copula approach perform worst according to the non-localising threshold-weighted scores, regardless of whether the energy score, variogram score or IMS is considered. However, when evaluated using a localising threshold-weighted score, or a vertically re-scaled score, these two approaches typically outperform the other post-processing methods. This is particularly the case for the energy score, even though the raw ensemble and the comonotonic approaches perform worst with respect to the unweighted energy score. 

\bigskip

\begin{figure}[b!]
    \centering
    \includegraphics[width = 0.48\linewidth]{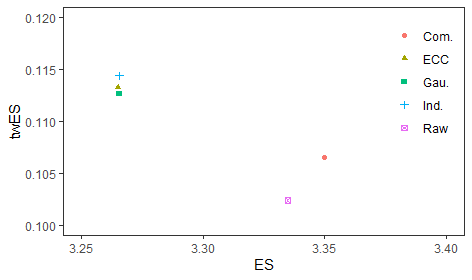}
    \caption{The energy score against the localising threshold-weighted energy score for the five multivariate forecast approaches.}
    \label{fig:MO_twESvsES}
\end{figure}

This highlights that although one forecast strategy might result in the best overall forecast performance, this may not be the preferred approach when interest is on high-impact events. To reinforce this, Figure \ref{fig:MO_twESvsES} displays the energy score against the (localising) threshold-weighted energy score for the five forecasting approaches. By analysing forecast performance with respect to multiple objectives, we can clearly see the trade-offs between the different approaches: the independence copula, ECC, and Gaussian copula generate forecasts that outperform the COSMO-E ensemble when evaluated using the energy score, but this comes at the expense of forecast accuracy when interest is on high-impact events.

\bigskip

Consider now employing a weight function in these weighted scores that is equal to one if the total precipitation over all three days exceeds 75mm, and is equal to zero otherwise - this is labelled ``total exceedance" in Table \ref{tab:mvScores}. In this case, the weight function acknowledges that flooding could result from moderate precipitation on consecutive days, or extreme precipitation on just a single day. The corresponding scores are displayed in Table \ref{tab:mvScores}. Since the simulation study in the previous section suggested that the non-localising variant of the threshold-weighted scores was not effective for this weight function, only the localising versions have been applied.

\bigskip

The independence copula, ECC, and Gaussian copula approaches generate the most accurate forecasts for these events, while the comonotonic copula and particularly the raw ensemble forecast perform comparatively poorly. This weight function depends less on the multivariate dependence structure than the previous weight function, and hence the results are more similar to those when an unweighted scoring rule is used to evaluate forecast performance.

\begin{table}[t]
    \centering
    \begin{tabular}{| c | c c c c c |} 
         \hline
         \multicolumn{6}{| c |}{Unweighted} \\
         \hline
         \multicolumn{1}{| c }{} & Raw & Ind. & ECC & Com. & Gau. \\
         \hline
         ES & 3.33 (0.03) & \textbf{3.27 (0.03)} & \textbf{3.27 (0.03)} & 3.35 (0.03) & \textbf{3.27 (0.03)} \\
         VS & \textbf{8.94 (0.08)} & 9.11 (0.08) & 9.12 (0.08) & 9.88 (0.09) & 9.11 (0.08) \\
         IMS & 26.8 (0.1) & \textbf{26.0 (0.1)} & \textbf{26.0 (0.1)} & 26.9 (0.1) & \textbf{26.0 (0.1)} \\
         \hline
         \multicolumn{6}{c}{ } \\
         \hline
         \multicolumn{6}{| c |}{Successive exceedance: $ w(z) = \one\{z_{1} \geq 25, z_{2} \geq 25, z_{3} \geq 25\} $ }\\
         \hline
         \multicolumn{1}{| c }{} & Raw & Ind. & ECC & Com. & Gau. \\
         \hline
         twES (non) & 0.733 (0.018) & \textbf{0.681 (0.017)} & \textbf{0.681 (0.017)} & 0.686 (0.018) & \textbf{0.681 (0.017)} \\
         twES (loc) & \textbf{0.102 (0.011)} & 0.114 (0.012) & 0.113 (0.012) & 0.107 (0.010) & 0.113 (0.012) \\
         vrES & \textbf{0.141 (0.014)} & 0.162 (0.017) & 0.161 (0.017) & 0.149 (0.014) & 0.160 (0.017) \\
         \hline
         twVS (non) & 3.26 (0.08) & \textbf{3.01 (0.08)} & \textbf{3.01 (0.08)} & 3.04 (0.08) & \textbf{3.01 (0.08)} \\
         twVS (loc) & 0.330 (0.041) & 0.333 (0.042) & 0.333 (0.042) & \textbf{0.318 (0.035)} & 0.328 (0.042) \\
         \hline
         twIMS (non) & 4.21 (0.07) & \textbf{4.05 (0.07)} & \textbf{4.05 (0.07)} & 4.07 (0.07) & \textbf{4.05 (0.07)} \\    
         twIMS (loc) & \textbf{0.169 (0.016)} & 0.179 (0.018) & 0.178 (0.018) & 0.175 (0.016) & 0.178 (0.018) \\    
         vrIMS & 0.123 (0.012) & \textbf{0.122 (0.012)} & \textbf{0.122 (0.012)} & 0.125 (0.012) & \textbf{0.122 (0.012)} \\    
         \hline
         \multicolumn{6}{c}{ } \\
         \hline
         \multicolumn{6}{| c |}{Total exceedance: $ w(z) = \one\{z_{1} + z_{2} + z_{3} \geq 75\} $} \\
         \hline
         \multicolumn{1}{| c }{} & Raw & Ind. & ECC & Com. & Gau. \\
         \hline
         twES (loc) & 0.674 (0.028) & 0.601 (0.027) & \textbf{0.600 (0.027)} & 0.602 (0.026) & 0.602 (0.027) \\
         vrES & 0.589 (0.025) & \textbf{0.535 (0.024)} & \textbf{0.535 (0.024)} & 0.540 (0.023) & 0.537 (0.024) \\
         \hline
         twVS (loc) & 2.45 (0.13) & \textbf{2.11 (0.11)} & \textbf{2.11 (0.11)} & 2.14 (0.11) & \textbf{2.11 (0.11)} \\
         \hline
         twIMS (loc) & 0.906 (0.036) & \textbf{0.865 (0.037)} & 0.866 (0.037) & 0.876 (0.036) & 0.867 (0.037) \\
         vrIMS & 0.621 (0.026) & \textbf{0.609 (0.026)} & \textbf{0.609 (0.026)} & 0.616 (0.026) & \textbf{0.609 (0.026)} \\
         \hline
    \end{tabular}
    \caption{Unweighted and weighted multivariate scoring rules for each of the five prediction methods. The scores have been averaged across all locations and all forecast instances in the test data set. Standard errors for the scores are shown in brackets, and the best approach with regard to each score is shown in bold. For convenience, all variogram scores have been scaled by 10, and all inverse multiquadric scores by 100.}
    \label{tab:mvScores}
\end{table}

\section{Discussion}
\label{section:Discussion}

Evaluating forecasts with an emphasis on extreme events is an intrinsically challenging task \citep{Taillardat2019, Brehmer2019}. Nonetheless, weighted scoring rules have become the standard approach to do so. In this article, we have introduced and examined weighted scoring rules that can be applied to multivariate forecasts. We contend that high-impact events often result from the interaction of several features, and the weighted multivariate scoring rules proposed herein therefore allow for a more thorough evaluation of forecasts with regards to high-impact events. 

\bigskip

The weighted multivariate scoring rules developed here have been constructed by exploiting existing theory on conditionally negative definite kernels and the associated kernel score framework. In particular, it is shown that the well-known threshold-weighted continuous ranked probability score is a kernel score. The effect of this is two-fold: firstly, it permits forecasts in the form of a finite ensemble to be evaluated easily using the threshold-weighted CRPS; secondly, the threshold-weighted CRPS can be generalised for use with alternative kernels, thereby producing a broader class of threshold-weighted kernel scores. 

\bigskip

In addition to this, well-known results on negative definite kernels have been leveraged in order to introduce a novel approach to weighting scoring rules. It is shown that this is equivalent to the threshold-weighting in particular circumstances, but in general extends the existing armory of weighted scores. Like the threshold-weighted kernel scores, these vertically re-scaled kernel scores also fall into the kernel score framework, making them applicable in a range of situations.

\bigskip

We explore these weighted scoring rules in the context of multivariate forecast evaluation, and compare them to alternative weighted scores previously proposed by \cite{Holzmann2017}. The energy score and the variogram score are the two most popular scoring rules when assessing multivariate forecasts, and both fall into the kernel score framework.  Additionally, we consider the inverse multiquadric score, a new kernel score for both univariate and multivariate outcomes based on a bounded kernel which is competitive with respect to discrimination ability and has the advantage of being strictly proper with respect to all probability measures on $\mathbb{R}^d$. We introduce various weighted energy scores, variogram scores, and inverse multiquadric scores that can emphasise particular outcomes when evaluating multivariate forecasts. We discuss possible ways that these kernel scores could be tuned in order to achieve this, and analyse the performance of the resulting scores in an application to simulated data in Section \ref{section:wES}. Although highly idealised, this study clearly demonstrates the utility of these weighted scores, highlighting the additional information they provide relative to the unweighted scores. 

\bigskip

It is then demonstrated how these weighted multivariate scoring rules could be applied in practice. In particular, several weighted multivariate scores are applied to MeteoSwiss daily precipitation accumulation forecasts, with a particular focus on events that could lead to flooding. Multivariate statistical post-processing methods are compared when forecasting these events, using both weighted and unweighted scores. Importantly, the conclusions drawn from the weighted scoring rules do not always coincide with those drawn from the unweighted scores, meaning the strategy that generates the most accurate forecasts is not necessarily the best when the goal is to predict flooding events. The weighted scoring rules can discriminate well between forecast distributions that exhibit contrasting behaviour in particular regions of the outcome space, something that unweighted scoring rules are not capable of.

\bigskip

Choosing a kernel that is relevant for a particular problem provides an extremely flexible approach to evaluate forecast performance, as illustrated by the fact that the variogram score is itself a kernel score. To this end, we believe that kernel scores have been underappreciated in the field of forecast verification. Kernels are used widely in several areas of mathematics and machine learning, and kernel scores could be employed to evaluate predictions made in a wide range of circumstances, including those for which established forecast verification tools do not currently exist. 

\bigskip

Results are presented here for kernel scores that employ a fairly simple weight or chaining function, and future applications of these scores could consider more elaborate weights. Furthermore, we expect that a more comprehensive analysis of forecast performance for multivariate outcomes will also inspire new developments in post-processing methodology with the goal of avoiding effects such as improving overall forecast performance at the cost of deteriorating accuracy when considering high-impact events.


\section*{Acknowledgements}

The authors are grateful to the Swiss Federal Office of Meteorology and Climatology (Meteoswiss) and the Oeschger Centre for Climate Change Research for financially supporting this work, and to Olivia Martius, Pascal Horton, Jonas Bhend, Lionel Moret, Mark Liniger and Jos{\'e} Carlos Araujo Acu{\~n}a for fruitful discussions during the preparation of this manuscript. Jonas Bhend is also thanked for providing the data that was used in the case study.

\bibliographystyle{apalike}
\bibliography{references}

\section*{Appendix}
\label{section:Appendix}

\subsection*{Proof of Theorem \ref{theorem:Proper}}

If $\rho$ is measurable and negative definite, then the result is a restatement of \citet[][Theorem 1.1]{Steinwart2019}. 

\bigskip

If $\rho$ is a measurable c.n.d.~kernel on $\mathcal{X}$ and $x_{0} \in \mathcal{X}$, then by \citet[][Lemma 2.1]{Berg1984}, the kernel
\begin{equation}\label{eq:Berg}
    k(x, x^{\prime}) = \rho(x, x^{\prime}) + \rho(x_{0}, x_{0}) - \rho(x, x_{0}) - \rho(x^{\prime}, x_{0})
\end{equation}
is negative definite, and also measurable, since $ \rho $ is measurable. Assuming all the relevant integrals exist and are finite, it is a straightforward calculation to show that $S_{k}(P, y) = S_{\rho}(P, y) $. By \citet[][Theorem 1.1]{Steinwart2019}, $S_{k}$ is a proper scoring rule with respect to the class of all probability measures $ P \in \mathfrak{M} $ such that $ \mathbb{E}_{P}[ \sqrt{-k(X, X)}] < \infty$. Note that $ -k(x, x) $ is non-negative, since $ - k$ is positive definite. In particular, $ S_{k} $ is a proper scoring rule with respect to the smaller class of all probability measures $ P \in \mathfrak{M} $ such that $ \mathbb{E}_{P}[-k(X, X)] < \infty $. 

\bigskip

We will now show that when $ \rho(x, x) = 0 $ for all $ x \in \mathcal{X} $, this class of measures coincides with $ \mathcal{M}_{\rho} $. We have that 
$$ k(x, x) = \rho(x, x) + \rho(x_{0}, x_{0}) - 2\rho(x, x_{0}), $$
and hence $ \rho(x, x) = 0 $ for all $ x \in \mathcal{X} $ implies that $ \rho $ is non-negative and thus a semi-metric. In this case, \citet[][Proposition 20]{Sejdinovic2013} state that $ P \in \mathcal{M}_{\rho} $ if and only if $ \mathbb{E}_{P}[-k(X, X)] < \infty $. Although this proposition is stated in the specific situation that $ \mathcal{A} $ is the Borel $ \sigma $-algebra with respect to a topology on $ \mathcal{X}$, this is not needed for the arguments in the proof, so the proposition holds for any measurable space $ (\mathcal{X}, \mathcal{A}) $. 

\bigskip

It follows that all integrals in Equation \ref{eq:KernelS} are finite for $ P \in \mathcal{M}_{\rho} $. This is the case, since, firstly, $ \mathbb{E}_{P}[\rho(X, x_{0})] < \infty $ for some $ x_{0} \in \mathcal{X} $ is equivalent to $ \mathbb{E}_{P}[\rho(X, x_{0})] < \infty $ for \emph{all} $x_{0} \in \mathcal{X}$, secondly, Equation \ref{eq:Berg} holds, and finally, $ -k(x, x^{\prime}) \leq \sqrt{-k(x, x)}\sqrt{-k(x^{\prime}, x^{\prime})} $.

\subsection*{Proof of Proposition \ref{prop:twcrps}}
\label{section:App_twCRPS}

Let $ \nu $ be a locally finite Borel measure on $ \R $, and let $ v: \R \to \R $ be a function such that $ v(x) - v(x^{\prime}) = \nu([x^{\prime}, x)) $ for any points $ x, x^{\prime} \in \R $. Then, 
\begin{equation*}
    \rho(x, x^{\prime}) = \vert v(x) - v(x^{\prime}) \vert = \int_{\R} \left(\one\{x^{\prime}\leq u < x\} + \one\{x\leq u < x^{\prime}\}\right) \dd \nu(u).
\end{equation*}   
If $ X, X^{\prime} $ are independent random variables distributed according to $ P, Q \in \mathcal{M}_{\rho} $ respectively, with associated distribution functions $ F $ and $ G $, then by the Fubini-Tonelli theorem,
\begin{equation*}
\begin{split}
    \mathbb{E}_{F,G} \vert v(X) - v(X^{\prime}) \vert & = \int_{\R}  \left( \mathbb{E}_{F,G}\left[ \one\{X^{\prime} \leq u < X\} \right] + \mathbb{E}_{F,G}\left[ \one\{X \leq u < X^{\prime} \}\right] \right) \dd \nu(u) \\
    & = \int_{\R}  G(u) \left[ 1 - F(u) \right]  + F(u) \left[ 1 - G(u) \right] \dd \nu(u). \\
\end{split}
\end{equation*} 
From this, it is straightforward to verify that if $ X, X^{\prime} \sim F $ are independent, then
\begin{equation*}
\begin{split}
  \mathbb{E}_{F} \vert v(X) - v(y) \vert  - \frac{1}{2} \mathbb{E}_{F} \vert v(X) - v(X^{\prime}) \vert = \int_{\R} \left(F(u) - \one\{y \leq u\} \right)^2 \dd \nu(u) = {\rm twCRPS}(F, y; \nu).
\end{split}
\end{equation*} 
Since $ \rho(x, x^{\prime}) $ is a c.n.d.~kernel, this proves that the threshold-weighted CRPS is a kernel score. 

\bigskip

To complete the proof of Proposition \ref{prop:twcrps}, it remains to show that this representation of the threshold-weighted CRPS is equivalent to a quantile scoring rule integrated over all $ \alpha \in (0, 1) $. The general form of a quantile scoring rule is
$$ QS_{v, \alpha}(F, y) = (\one\{F^{-1}(\alpha) \geq y\} - \alpha)(v(F^{-1}(\alpha)) - v(y)) $$
for some increasing function $ v $, where $ F^{-1} $ is the generalised inverse of $ F $. Integrating this with respect to $ \alpha $ gives
\begin{equation*}
\begin{split}
    2 \int_{(0, 1)} & (\one\{F^{-1}(\alpha) \geq y\} - \alpha)(v(F^{-1}(\alpha)) - v(y)) \dd \alpha \\
    = & \int_{(0, 1)} (2\one\{F^{-1}(\alpha) \geq y\} - 1)(v(F^{-1}(\alpha)) - v(y)) \dd \alpha - \int_{(0, 1)} (2\alpha - 1)(v(F^{-1}(\alpha)) - v(y)) \dd \alpha \\
    = & \int_{(0, 1)} |v(F^{-1}(\alpha)) - v(y)| \dd \alpha - \int_{\R} (2F(x) - 1)(v(x) - v(y)) \dd F(x), \\
    = & \int_{\R} |v(x) - v(y)| \dd F(x) - \int_{\R} \int_{\R} (2\one\{x^{\prime} \leq x\} - 1)(v(x) - v(y)) \dd F(x^{\prime})\dd F(x).
\end{split}
\end{equation*}
The first term is equivalent to $ \mathbb{E}_{F}|v(X) - v(y)| $ with $ X \sim F $, while the latter can be rewritten as
\begin{equation*}
\begin{split}
    I_{2} & = \mathbb{E}_{F}\left[ (2 \one\{X^{\prime} \leq X \} - 1)(v(X) - v(y)) \right] \\
    & = \mathbb{E}_{F}\left[ (2 \one\{X^{\prime} \leq X \} - 1)(v(X) - v(X^{\prime})) \right] + \mathbb{E}_{F}\left[ (2 \one\{X^{\prime} \leq X \} - 1)(v(X^{\prime}) - v(y)) \right] \\
    & = \mathbb{E}_{F} |v(X) - v(X^{\prime})| - \mathbb{E}_{F} \left[ (2 \one\{X < X^{\prime}\} - 1)(v(X^{\prime}) - v(y)) \right] \\
    & = \mathbb{E}_{F} |v(X) - v(X^{\prime})| - I_{2},
\end{split}
\end{equation*}
where $ X,X^{\prime} \sim F $ are independent. Rearranging gives $ I_{2} = \mathbb{E}_{F} |v(X) - v(X^{\prime})|/2 $, which in turn yields 
\begin{equation*}
    2 \int_{(0, 1)} (\one\{F^{-1}(\alpha) \geq y\} - \alpha)(v(F^{-1}(\alpha)) - v(y)) \dd \alpha = \mathbb{E}_{F}|v(X) - v(y)| - \frac{1}{2} \mathbb{E}_{F} |v(X) - v(X^{\prime})|,
\end{equation*}
as desired.

\subsection*{Proof of Proposition \ref{prop:owcrps}}
\label{section:App_owCRPS}

Let $ \mathcal{M} $ denote the set of Borel probability measures on $ \mathcal{X} = \R $ with finite first moment. Let $ F \in \mathcal{M}  $ such that $ \mathbb{E}_{F}[w(X)] > 0 $, and define $ F_{w} $ as in Equation \ref{eq:weightedG}. Since $ F \in \mathcal{M} $, we also have that $ F_{w} \in \mathcal{M} $. Then, the outcome-weighted CRPS with weight function $ w $ can be written as
\begin{equation*}
\begin{split}
    \mathrm{owCRPS}(F, y; w) & = w(y)\mathrm{CRPS}(F_{w}, y) \\
    & = \mathbb{E}_{F_{w}} |Z - y| w(y) - \frac{1}{2} \mathbb{E}_{F_{w}}  |Z - Z^{\prime}|  w(y), \\
    & = \frac{1}{C_{w}(F)} \mathbb{E}_{F} \left[ |X - y|w(X)w(y) \right] - \frac{1}{2C_{w}(F)^{2}} \mathbb{E}_{F} \left[ |X - X^{\prime}|w(X)w(X^{\prime})w(y) \right],
\end{split}
\end{equation*}
where $ Z, Z^{\prime} \sim F_{w} $ and $ X, X^{\prime} \sim F $ are independent.

\subsection*{Proof of Proposition \ref{prop:twks}}

Let $ \rho $ be a c.n.d.~kernel on $ \mathcal{X} $, let $ v: \mathcal{X} \to \mathcal{X} $ be a measurable function, and let $ \tilde{\rho}(x, x^{\prime}) = \rho(v(x), v(x^{\prime})) $. Suppose that $ S_{\rho} $ is a strictly proper scoring rule with respect to $ \mathcal{M}_{\rho} $. 

\bigskip

Firstly, assume that $ v $ is injective. Let $ \tilde{P}, \tilde{Q} \in \mathcal{M}_{\tilde{\rho}} $. We wish to show that $ d_{\tilde{\rho}}(\tilde{P}, \tilde{Q}) = 0 $ implies $ \tilde{P} = \tilde{Q} $. To do so, define $ P $ and $ Q $ as the push-forward of $ v $ under $ \tilde{P} $ and $ \tilde{Q} $, respectively, i.e.~$ P(A) = \tilde{P}(v^{-1}(A)) $ and $ Q(A) = \tilde{Q}(v^{-1}(A)) $ for all $ A \in \mathcal{A} $. Note that, since $ v $ is injective, it is also bi-measurable. Additionally, for $ \tilde{P} \in \mathcal{M}_{\tilde{\rho}} $, we have $ \mathbb{E}_{\tilde{P}}[\rho(v(X), v(x_{0})] < \infty $ for some $ x_{0} \in \mathcal{X} $. Letting $ x_{1} = v(x_{0}) $, 
$$ \mathbb{E}_{\tilde{P}}[\rho(v(X), v(x_{0})] = \mathbb{E}_{P}[\rho(X, x_{1})] < \infty, $$
and hence $ P, Q \in \mathcal{M}_{\rho} $.

\bigskip

The divergence function of the kernel score associated with $ \tilde{\rho} $ can then be written as
\begin{equation*}
\begin{split}
    d_{\tilde{\rho}}(\tilde{P}, \tilde{Q}) & = \mathbb{E}_{P, Q} \left[ \rho(X, Y) \right] - \frac{1}{2} \mathbb{E}_{P} \left[ \rho(X, X^{\prime}) \right] - \frac{1}{2} \mathbb{E}_{Q} \left[ \rho(Y, Y^{\prime}) \right] \\
    & = d_{\rho}(P, Q),
\end{split}
\end{equation*}
where $ X, X^{\prime} \sim P $ and $ Y, Y^{\prime} \sim Q $ are independent. Since $ S_{\rho} $ is strictly proper with respect to $ \mathcal{M}_{\rho} $, $ d_{\rho}(P, Q) = 0 $ implies $ P = Q $. Then, for any $ A \in \mathcal{A} $, we have that
$$ \tilde{P}(A) = \tilde{P}(v^{-1}(v(A))) = P(B) = Q(B) = \tilde{Q}(v^{-1}(v(A))) = \tilde{Q}(A), $$
where $ B = v(A) \in \mathcal{A} $, and $ v^{-1}(v(A)) = A $ holds for all $ A \in \mathcal{A} $ since $ v $ is injective. Hence, $ d_{\tilde{\rho}}(\tilde{P}, \tilde{Q}) = 0 $ implies that $ \tilde{P} = \tilde{Q} $.

\bigskip

Conversely, assume that $\mathrm{tw}S_{\rho} $ is strictly proper with respect to $ \mathcal{M}_{\tilde{\rho}} $. If $ v $ is not injective, then there exist distinct $ z, z^{\prime} \in \mathcal{X} $ such that $ v(z) = v(z^{\prime}) $. Letting $ \tilde{P} = \delta_{z} $ and $ \tilde{Q} = \delta_{z^{\prime}} $ be Dirac measures at $ z $ and $ z^{\prime} $, respectively, we have
\begin{equation*}
\begin{split}
    d_{\tilde{\rho}}(\tilde{P}, \tilde{Q}) & = \rho(v(z), v(z^{\prime})) - \frac{1}{2} \rho(v(z), v(z)) - \frac{1}{2} \rho(v(z^{\prime}), v(z^{\prime})), \\
    & = \rho(v(z), v(z)) - \frac{1}{2} \rho(v(z), v(z)) - \frac{1}{2} \rho(v(z), v(z)) = 0.
\end{split}
\end{equation*}
That is, there exist distinct $ \tilde{P}, \tilde{Q} \in \mathcal{M}_{\tilde{\rho}} $ such that $ d_{\tilde{\rho}}(\tilde{P}, \tilde{Q}) = 0 $, which contradicts the assumption that $\mathrm{tw}S_{\rho} $ is strictly proper with respect to $ \mathcal{M}_{\tilde{\rho}} $. Hence, if the kernel score associated with $ \rho $ is strictly proper with respect to $ \mathcal{M}_{\rho} $, then the kernel score associated with $ \tilde{\rho} $ is strictly proper with respect to $\mathcal{M}_{\tilde{\rho}} $ if and only if the chaining function $ v $ is injective. Similar arguments can be used in the case that $ S_{\rho} $ is strictly proper with respect to $ \mathcal{M}^{\rho} $.

\subsection*{Proof of Proposition \ref{prop:twks_lp}}

Let $ \rho $ be a c.n.d.~kernel on $ \mathcal{X}$ such that $ \rho(x, x) = 0 $ for all $ x \in \mathcal{X} $, let $ w $ be a weight function, and let $ v: \mathcal{X} \to \mathcal{X} $ be measurable. Set $ \tilde{\rho}(x, x^{\prime}) = \rho(v(x), v(x^{\prime})) $. For $ P \in \mathcal{M}_{\tilde{\rho}} $, define $P_{0} = P(\cdot \cap \{w = 0\}) $ and $P_{+} = P(\cdot \cap \{w > 0\}) $, where $ \{w > 0\} = \{x \in \mathcal{X} | w(x) > 0\} $ and $ \{w = 0\} = \{x \in \mathcal{X} | w(x) = 0\} $. Since $ \mathcal{X} $ can be partitioned into the union of $ \{ w > 0 \} $ and $ \{ w = 0 \} $, the threshold-weighted kernel score with kernel $ \rho $ and chaining function $ v $ can be decomposed as
\begin{equation}\label{eq:twks_lp}
\begin{split}
    \mathrm{tw}S_{\rho}(P, y; v) = & \int_{\mathcal{X}} \rho(v(x), v(y)) \dd P_{0}(x) - \frac{1}{2} \int_{\mathcal{X}} \int_{\mathcal{X}} \rho(v(x), v(x^{\prime})) \dd P_{0}(x) \dd P_{0}(x^{\prime}) \\
    & + \int_{\mathcal{X}} \rho(v(x), v(y)) \dd P_{+}(x) - \frac{1}{2} \int_{\mathcal{X}} \int_{\mathcal{X}} \rho(v(x), v(x^{\prime})) \dd P_{+}(x) \dd P_{+}(x^{\prime}) \\
    & - \int_{\mathcal{X}} \int_{\mathcal{X}} \rho(v(x), v(x^{\prime})) \dd P_{0}(x) \dd P_{+}(x^{\prime}).
\end{split}
\end{equation}

\bigskip

Suppose that $ \rho(v(z), v(\cdot)) = \rho(v(z^{\prime}), v(\cdot)) $ for all $ z, z^{\prime} \in \{w = 0\} $. Note that this implies that $ \rho(v(z), v(z^{\prime})) = 0 $ for all $ z, z^{\prime} \in \{w = 0\} $. In this case, $ \mathrm{tw}S_{\rho}(P, y; v) $ simplifies to
\begin{equation*}
\begin{split}
    \mathrm{tw}S_{\rho}(P, y; v) = & \rho(v(x_{0}), v(y)) P(\{w = 0\}) \\
    & + \int_{\mathcal{X}} \rho(v(x), v(y)) \dd P_{+}(x) - \frac{1}{2} \int_{\mathcal{X}} \int_{\mathcal{X}} \rho(v(x), v(x^{\prime})) \dd P_{+}(x) \dd P_{+}(x^{\prime}) \\
    & - P(\{w = 0\}) \int_{\mathcal{X}} \rho(v(x), v(x_{0})) \dd P_{+}(x),
\end{split}
\end{equation*}
where $ x_{0} $ is an arbitrary point in $ \{w = 0\} $. Since any $ P, Q \in \mathcal{M}_{\tilde{\rho}} $ are probability measures, if $ P_{+} = Q_{+} $, then $ P(\{w = 0\}) = Q(\{w = 0\}) $ and it becomes clear that $ \mathrm{tw}S_{\rho}(P, y; v) = \mathrm{tw}S_{\rho}(Q, y; v) $ for all $ y \in \mathcal{X} $. Hence, the threshold-weighted kernel score with such a kernel and chaining function is localising with respect to $ w $.

\bigskip

Suppose now that $ \mathrm{tw}S_{\rho} $ is localising with respect to $ w $. Consider 
$$ P = \frac{1}{2} \delta_{z} + \frac{1}{2} \delta_{x}, \hspace{2cm} Q = \frac{1}{2} \delta_{z^{\prime}} + \frac{1}{2} \delta_{x}, $$
where $ x \in \{w > 0 \} $ and $ z, z^{\prime} \in \{w = 0\} $. Note that $ P, Q \in \mathcal{M}_{\tilde{\rho}} $ and $ P(\cdot \cap \{w > 0\}) = Q(\cdot \cap \{w > 0 \}) $. Since $ \mathrm{tw}S_{\rho} $ is localising with respect to $ w $, we have that $ \mathrm{tw}S_{\rho}(P, y; v) - \mathrm{tw}S_{\rho}(Q, y; v) = 0 $ for all $ y \in \mathcal{X} $. From Equation \ref{eq:twks_lp}, this means that 
\begin{equation}\label{eq:twks_lp_scoredif}
\begin{split}
    \mathrm{tw}S_{\rho}(P, y; v) - \mathrm{tw}S_{\rho}(Q, y; v) & = \frac{1}{2} \rho(v(z), v(y)) - \frac{1}{4} \rho(v(z), v(x)) - \frac{1}{2} \rho(v(z^{\prime}), v(y)) + \frac{1}{4} \rho(v(z^{\prime}), v(x)), \\
    & = 0,
\end{split}
\end{equation}
for all $ z, z^{\prime} \in \{w = 0\} $, $ x \in \{w > 0\} $, and $ y \in \mathcal{X} $.

\bigskip

Since this holds for all $ y \in \mathcal{X} $, we can substitute $ y = x $ to yield $ \rho(v(z), v(x)) = \rho(v(z^{\prime}), v(x)) $ for all $ x \in \{w > 0\} $, $ z, z^{\prime} \in \{w = 0\} $. Using this, and substituting $ y = z^{\prime} $ into Equation \ref{eq:twks_lp_scoredif}, we additionally have $ \rho(v(z), v(z^{\prime})) = 0 $ for all $ z, z^{\prime} \in \{w = 0 \} $. Hence, if $ \mathrm{tw}S_{\rho} $ is localising with respect to $ w $, then $ \rho(v(z), v(x)) = \rho(v(z^{\prime}), v(x)) $ for all $ z, z^{\prime} \in \{w = 0 \} $ and $ x \in \mathcal{X} $.

\subsection*{Proof of Proposition \ref{prop:twks_slp}}

Let $ \rho $ be a c.n.d.~kernel, let $ v: \mathcal{X} \to \mathcal{X} $ be measurable, and define $ \tilde{\rho}(x, x^{\prime}) = \rho(v(x), v(x^{\prime})) $. Suppose that $ S_{\rho} $ is strictly proper with respect to $ \mathcal{M}_{\rho} $.

\bigskip

Firstly, assume that the restriction of $ v $ to $ \{w > 0\} $ is injective; that is, $ v(z) \neq v(z^{\prime}) $ for all $ z, z^{\prime} \in \{w > 0\} $. Let $ \tilde{P}, \tilde{Q} \in \mathcal{M}_{\tilde{\rho}} $. We wish to show that $ d_{\tilde{\rho}}(\tilde{P}, \tilde{Q}) = 0 $ implies $ \tilde{P}(\cdot \cap \{w > 0\}) = \tilde{Q}(\cdot \cap \{w > 0\}) $. Define $ P $ and $ Q $ as the push-forward of $ v $ under $ \tilde{P} $ and $ \tilde{Q} $, respectively, i.e.~$ P(A) = \tilde{P}(v^{-1}(A)) $ and $ Q(A) = \tilde{Q}(v^{-1}(A)) $ for all $ A \in \mathcal{A} $. From the proof of Proposition \ref{prop:twks}, we know that $ P, Q \in \mathcal{M}_{\rho} $. Also from this proof, we have that  $ d_{\tilde{\rho}}(\tilde{P}, \tilde{Q}) = 0 $ implies that $ P = Q $. Then, for any $ A \in \mathcal{A} $, we have that
$$ \tilde{P}(A \cap \{w > 0\}) = \tilde{P}(v^{-1}(v(A \cap \{w > 0\}))) = P(B) = Q(B) = \tilde{Q}(v^{-1}(v(A \cap \{w > 0\}))) = \tilde{Q}(A), $$
where $ B = v(A \cap \{w > 0\}) \in \mathcal{A} $, and $ v^{-1}(v(A \cap \{w > 0\})) = A \cap \{w > 0\} $ for all $ A \in \mathcal{A} $ since the restriction of $ v $ to $ \{ w > 0\} $ is injective. Hence, $ d_{\tilde{\rho}}(\tilde{P}, \tilde{Q}) = 0 $ implies that $ P = Q $, which in turn implies that $ \tilde{P}(\cdot \cap \{w > 0\}) = \tilde{Q}(\cdot \cap \{w > 0\}) $. The threshold-weighted kernel score with this loss function is therefore strictly locally proper with respect to $ w $ and $ \mathcal{M}_{\tilde{\rho}} $.

\bigskip

Conversely, assume that $ \mathrm{tw}S_{\rho} $ is strictly locally proper with respect to $ w $ and $ \mathcal{M}_{\tilde{\rho}} $. If the restriction of $ v $ to $ \{w > 0\}$ is not injective, then there exist distinct $ z, z^{\prime} \in \{w > 0\} $ such that $ v(z) = v(z^{\prime}) $. If $ \tilde{P} = \delta_{z} $ and $ \tilde{Q} = \delta_{z^{\prime}} $ are Dirac measures at $ z $ and $ z^{\prime} $, respectively, then $ d_{\tilde{\rho}}(\tilde{P}, \tilde{Q}) = 0 $, even though $ \tilde{P}(\cdot \cap \{w > 0\}) = \tilde{Q}(\cdot \cap \{w > 0 \}) $. This is a contradiction, and hence if $ \mathrm{tw}S_{\rho} $ is strictly locally proper with respect to $ w $ and $ \mathcal{M}_{\tilde{\rho}} $, then the restriction of $ v $ to $ \{ w > 0 \} $ must be injective. Similar arguments can be used in the case that $ \mathrm{tw}S_{\rho} $ is strictly proper with respect to $ \mathcal{M}^{\tilde{\rho}} $.

\subsection*{Proof of Propositions \ref{prop:nsks} and \ref{prop:nsks_lp}}

Let $-\rho$ be strictly integrally positive definite with respect to the maximal possible set of signed measures on $\{w > 0\}$ in the sense of \citet[Definition 2.1]{Steinwart2019}, let $ w>0$ be a weight function, and let $ \check{\rho}(x, x^{\prime}) = \rho(x, x^{\prime})w(x)w(x^{\prime}) $. 

\bigskip

Let $ P, Q \in \mathcal{M}^{\rho} $ and let $\tilde P, \tilde Q$ be the measures on $\{w > 0\}$ that are absolutely continuous with respect to $P(\cdot \cap\{w > 0\})$ and $Q(\cdot \cap\{w > 0\})$, respectively, and density $w$. Note that $\tilde P, \tilde Q$ are finite measures on $\{w > 0\}$ but not necessarily probability measures. For the divergence function corresponding to $ \mathrm{vr}S_{\rho}(\cdot, \cdot; w) $, we obtain that
\begin{align*}
   0= d_{\check{\rho}}(P, Q) &= \int_{\{w > 0\}} \int_{\{w > 0\}} \rho(x,y) \dd \tilde P(x)\dd\tilde Q(y) - \frac{1}{2} \int_{\{w > 0\}} \int_{\{w > 0\}} \rho(x,x') \dd \tilde P(x)\dd\tilde P(x') \\& \qquad- \frac{1}{2} \int_{\{w > 0\}} \int_{\{w > 0\}} \rho(y,y') \dd \tilde Q(y)\dd\tilde Q(y')
\end{align*}
implies that $\tilde P = \tilde Q$ due to $-\rho$ being integrally strictly positive definite on $\{w > 0\}$. This yields that $P(\cdot \cap\{w > 0\}) = Q(\cdot \cap\{w > 0\})$.

%
%
%
%
%

\subsection*{Proof of Proposition \ref{prop:nsks_twks}}

Let $ \rho $ be a c.n.d.~kernel on $ \mathcal{X} $ with $ \rho(x, x) = 0 $ for all $ x \in \mathcal{X} $, and let $ w $ be a weight function such that $ w(x) \in \{0, 1\} $ for all $ x \in \mathcal{X} $. Consider the chaining function $ v(x) = xw(x) + x_{0}(1 - w(x)), $ for $ x, x_{0} \in \mathcal{X} $. 

\bigskip

For this chaining function, we have that
\begin{equation*}
    \rho(v(x), v(x^{\prime})) = \rho(x, x^{\prime})w(x)w(x^{\prime}) + \rho(x, x_{0})w(x)(1 - w(x^{\prime})) + \rho(x^{\prime}, x_{0})w(x^{\prime})(1 - w(x)).
\end{equation*}

Let $ P \in \mathcal{M}_{\rho} $. Assuming all expectations are finite, substituting the above kernel into Equation \ref{eq:twks_def} and rearranging gives
\begin{equation*}
\begin{split}
    \mathrm{tw}S_{\rho}(P, y; v) = & \mathbb{E}_{P} \left[ \rho(X, y) w(X)w(y) \right] - \frac{1}{2} \mathbb{E}_{P} \left[\rho(X, X^{\prime}) w(X) w(X^{\prime}) \right] \\
    & + ( \mathbb{E}_{P} \left[ \rho(X, x_{0})w(X) \right] - \rho(y, x_{0})w(y) )(\mathbb{E}_{P}[w(X)] - w(y)),
\end{split}
\end{equation*}
which is the vertically re-scaled kernel score with kernel $ \rho $, weight $ w $, and centre $ x_{0} $, as given in Equation \ref{eq:nsks2}.

\subsection*{Gaussian copula implementation}

Copula analysis involves modelling a multivariate variable by first fitting distributions to each of the marginal variables, and then modelling the dependence between these marginal variables. In practice, this often requires sampling from the copula. If the copula is parametric, then this is typically performed randomly. However, it is sometimes beneficial to restrict the values of the copula that can be sampled. For example, by ensuring that the sampled points along each dimension must be equal to particular quantiles of the marginal distributions. That is, by restricting the domain of the copula from $[0, 1]^{d}$ to $\{ \tau_{1}, \dots, \tau_{M} \}^{d} $ for quantile levels $ \tau_{1}, \dots, \tau_{M} \in [0, 1] $, the canonical choice being   $ \tau_{i} = i/(M + 1) $ for $ i = 1, \dots, M $, which we have employed.

\bigskip

For example, in the field of statistical post-processing, using equidistant quantile levels typically results in a forecast ensemble that is more accurate than an ensemble generated at random from the copula \citep{Schefzik2013, Lerch2020}. As such, we introduce an approach that allows us to sample particular quantiles from the marginal distributions, and reorder them such that the dependence structure of the fitted copula is well-represented. 

\bigskip

In theory, this could be achieved by computing the likelihood implied by the copula for all possible permutations of these $ M $ quantiles in each dimension, and choosing the permutation associated with the highest likelihood. However, the number of such permutations is $ (M!)^{d-1} $, making this approach computationally infeasible unless both $ M $ and the dimension $ d $ are very small. 

\bigskip

Each permutation in the approach described in the previous paragraph is comprised of $ M $ combinations of the quantiles/quantile levels in the different dimensions. Instead, we calculate the likelihood of each individual combination of the quantile levels, reducing the number of evaluations to $ M^{d} $. That is, we evaluate the copula density at the $M^d$ points on the grid $\{ \tau_{1}, \dots, \tau_{M} \}^{d} $. A combination is sampled at random with probabilities that are proportional to the likelihood derived from the copula: combinations of the quantile levels that receive a high likelihood are more likely to be selected. Combinations that share the same quantile along any of the dimensions as the chosen combination are removed from consideration. This process is then repeated until we have $ M $ combinations. The independence copula and the comonotonic copula can both be interpreted within this general framework: for the independence copula, the probabilities corresponding to all combinations are the same, while for the comonotonic copula, the probability is one for the comonotonic combinations and zero otherwise.

\bigskip

\begin{figure}
    \centering
    \includegraphics[width=0.48\textwidth]{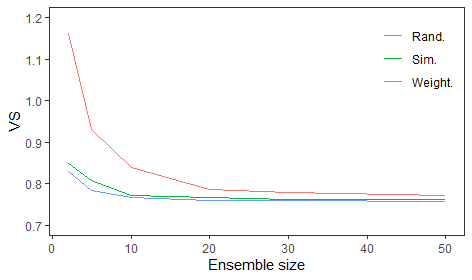}
    \caption{Variogram score against the ensemble size for three Gaussian copula-based post-processing methods applied to the MeteoSwiss COSMO-E ensemble forecasts of Section \ref{section:CaseStudy}. The three approaches either sample randomly from the copula (Rand.), simulate combinations of particular quantile levels with probabilities proportional to the copula likelihood (Sim.), or weight each possible combination using this likelihood (Weight.).}
    \label{fig:GauCop_VS}
\end{figure}

Although this approach does not consider every possible permutation of the quantiles, it provides a computationally achievable alternative. When applied to the Gaussian copula-based post-processing method in the case study of Section \ref{section:CaseStudy}, it is found to generate ensemble forecasts that are significantly more accurate than those constructed by sampling from the copula at random. Figure \ref{fig:GauCop_VS} shows the variogram score for these two approaches as a function of the ensemble size. The approach is particularly beneficial when the ensemble size is small (less than 20 members). For the results in Section \ref{section:CaseStudy}, we choose $ M $ to be the number of members of the COSMO-E ensemble forecast, but we could easily have applied this approach with a larger number of ensemble members, which should result in better forecast performance.

\bigskip

Although this appears to be more beneficial than simulating randomly from the copula, sampling particular combinations of the quantiles still neglects information provided by the copula. As an alternative, rather than sampling from the set of combinations with probabilities proportional to the likelihood, these likelihoods could be used to weight each possible combination. In the post-processing set up, this would result in an ensemble forecast of $ M^{d} $ members, each of which is assigned a weight that is proportional to the likelihood. Figure \ref{fig:GauCop_VS} also shows the variogram score for this approach, which offers slight improvements upon the previously described approach.

\end{document}